\def\beq#1{\begin{equation}\label{#1}}
\def\eeq{\end{equation}}
\def\beqa#1{\begin{eqnarray}\label{#1}}
\def\eeqa{\end{eqnarray}}

\def\fun#1#2{\lower3.6pt\vbox{\baselineskip0pt\lineskip.9pt
        \ialign{$\mathsurround=0pt#1\hfill##\hfil$\crcr#2\crcr\sim\crcr}}}

\def\xi{{{\bf x}^b}}

\documentclass
[onecolumn,aps,showpacs,showkeys,nofootinbib,superscriptaddress,amsmath]{revtex4}
\usepackage{epsfig}

\newcommand{\be}{\begin{equation}}
\newcommand{\ee}{\end{equation}}
\newcommand{\ba}{\begin{eqnarray}}
\newcommand{\ea}{\end{eqnarray}}

\begin{document}
\input{epsf.sty}

\title{Effects of time-varying $\beta$ in SNLS3 on constraining interacting dark energy models}

\author{Shuang Wang}
\email{wjysysgj@163.com}
\affiliation{Department of Physics, College of Sciences, Northeastern University, Shenyang 110004, China}

\author{Yong-Zhen Wang}
\email{w_avin@163.com}
\affiliation{Department of Physics, College of Sciences, Northeastern University, Shenyang 110004, China}

\author{Jia-Jia Geng}
\email{gengjiajia163@163.com}
\affiliation{Department of Physics, College of Sciences, Northeastern University, Shenyang 110004, China}

\author{Xin Zhang\footnote{Corresponding author}}
\email{zhangxin@mail.neu.edu.cn}
\affiliation{Department of Physics, College of Sciences, Northeastern University, Shenyang 110004, China}
\affiliation{Center for High Energy Physics, Peking University, Beijing 100080, China}

\begin{abstract}
It has been found that, for the Supernova Legacy Survey three-year (SNLS3) data,
there is strong evidence for the redshift-evolution of color-luminosity parameter $\beta$.
In this paper, adopting the $w$-cold-dark-matter ($w$CDM) model and considering
its interacting extensions (with three kinds of interaction between dark sectors),
we explore the evolution of $\beta$ and its effects on parameter estimation.
In addition to the SNLS3 data, we also use the latest Planck distance priors data,
the galaxy clustering (GC) data extracted from
sloan digital sky survey (SDSS) data release 7 (DR7) and baryon oscillation spectroscopic survey (BOSS),
as well as the direct measurement of Hubble constant $H_0$ from the Hubble Space Telescope (HST) observation.
We find that, for all the interacting dark energy (IDE) models,
adding a parameter of $\beta$ can reduce $\chi^2$ by $\sim$ 34,
indicating that a constant $\beta$ is ruled out at 5.8$\sigma$ confidence level (CL).
Furthermore, it is found that varying $\beta$ can significantly change the fitting results of various cosmological parameters:
for all the dark energy models considered in this paper,
varying $\beta$ yields a larger fractional CDM densities $\Omega_{c0}$ and a larger equation of state $w$;
on the other side, varying $\beta$ yields a smaller reduced Hubble constant $h$ for the $w$CDM model,
but has no impact on $h$ for the three IDE models.
This implies that there is a degeneracy between $h$ and coupling parameter $\gamma$.
Our work shows that the evolution of $\beta$ is insensitive to the interaction between dark sectors,
and then highlights the importance of considering $\beta$'s evolution in the cosmology fits.

\end{abstract}



\maketitle

\section{Introduction}

Cosmic acceleration is one of the biggest puzzles in modern cosmology
\cite{Riess98,spergel03,Tegmark04,Komatsu09,Percival10,Drinkwater10,Riess11}.
There are mainly two approaches to explain this extremely counterintuitive phenomenon:
dark energy (DE) \cite{quint,phantom,k,Chaplygin,ngcg,tachyonic,HDE,hessence,YMC,hscalar,cq,others1,others2,others3,others4}
and modified gravity (MG) \cite{SH,PR,DGP,GB,Galileon,FR,FT,FRT}.
For recent reviews, see \cite{CST,FTH,Linder,CK,Uzan,Tsujikawa,NO,LLWW,CFPS,YWBook}.

Cosmological observations are of essential importance to understanding cosmic acceleration,
and one of the most important observations is Type Ia supernovae (SNe Ia) \cite{Union,Constitution,Union2,Union2.1}.
In 2010, the Supernova Legacy Survey (SNLS) group released their three years data, i.e. SNLS3 dataset \cite{Guy10}.
Soon after, using this dataset,
Conley et al. \cite{Conley11} and Sullivan et al. \cite{Sullivan11}
presented the SN-only cosmological results and the joint cosmological constraints, respectively.
Unlike other supernova (SN) group, the SNLS team treated two important quantities,
stretch-luminosity parameter $\alpha$ and color-luminosity parameter $\beta$ of SNe Ia,
as free model parameters.

There are many factors that can lead to systematic uncertainties of SNe Ia.
One of the most important factors is the potential SN evolution,
i.e. the possibility for the redshift evolution of $\alpha$ and $\beta$.
The current studies show that $\alpha$ is consistent with a constant,
but the hints for the evolution of $\beta$ have been found in \cite{Astier06,Kessler09,Marriner11,Scolnic1,Scolnic2}.
For example, in \cite{Mohlabeng}, using a linear $\beta(z) = \beta_0 + \beta_1 z$,
Mohlabeng and Ralston studied the case of Union2.1 dataset
and found that $\beta$ deviates from a constant at 7$\sigma$ confidence levels (CL).
In \cite{WangWang}, Wang \& Wang found that, for the SNLS3 data,
$\beta$ increases significantly with $z$ at the 6$\sigma$ CL;
moreover, they proved that this conclusion is insensitive to the lightcurve fitter models,
or the functional form of $\beta(z)$ assumed \cite{WangWang}.
Therefore, the evolution of $\beta$ is a common phenomenon for various SN datasets,
and should be taken into account seriously.

It is very interesting to study the effects of a time-varying $\beta$ on parameter estimation.
In \cite{WangNew}, Wang, Li \& Zhang explored this issue
by considering the $\Lambda$-cold-dark-matter ($\Lambda$CDM) model, the $w$CDM model, and the Chevallier-Polarski-Linder (CPL) model.
Soon after, Wang, Geng, Hu \& Zhang \cite{WangNew2} studied the case of holographic dark energy (HDE) model,
which is a physically plausible DE candidate based on the holographic principle.
It is found that, for all these DE models, adding a parameter of $\beta$ can reduce $\chi^2_{min}$ by $\sim$ 36;
in addition, considering the evolution of $\beta$ is helpful in reducing the tension between SN and other cosmological observations.
It should be mentioned that,
in principle there is always an important possibility that DE directly interacts with CDM.
This factor was not considered in Refs. \cite{WangNew} and \cite{WangNew2}.
To do a comprehensive analysis on the effects of a time-varying $\beta$,
it is necessary to extend the corresponding discussions to the case of interacting dark energy (IDE) models.

In this paper, we explore the effects of a time-varying $\beta$ on the cosmological constraints of the IDE model.
Three kinds of interaction terms are taken into account.
In addition to the SNLS3 data,
we also use the Planck distance prior data \cite{WangWangCMB},
the galaxy clustering (GC) data from SDSS DR7 \cite{ChuangWang12} and BOSS \cite{Chuang13},
as well as the direct measurement of Hubble constant $H_0=73.8\pm 2.4 {\rm km/s/Mpc}$ from the Hubble Space Telescope (HST) observation \cite{Riess11}.

We describe our method in Sec.~II, present our results in Sec.~III, and conclude in Sec.~IV.
In this paper, we assume today's scale factor $a_{0}=1$, thus the redshift $z=a^{-1}-1$.
The subscript ``0'' always indicates the present value of the corresponding quantity,
and the natural units are used.

\section{Methodology}
\label{sec:method}

\subsection{Theoretical models}

In this paper, we consider a non-flat universe.
%which is filled with four major energy components: CDM, DE, radiation and baryons.
The Friedmann equation can be written as
\begin{equation}\label{F.e.}
    3M_{pl}^{2}H^{2}=\rho_{c}+\rho_{de}+\rho_{r}+\rho_{b}+\rho_{k},
\end{equation}
where $M_{pl}\equiv 1/\sqrt{8\pi G}$ is the reduced Planck mass,
$\rho_{c}$, $\rho_{de}$, $\rho_{r}$, $\rho_{b}$ and $\rho_{k}$
are the energy densities of CDM, DE, radiation, baryon and curvature, respectively.
The reduced Hubble parameter $E(z)\equiv H(z)/H_{0}$ satisfies
\begin{equation}\label{E}
\begin{split}
E^{2}=&\Omega_{c0}\frac{\rho_{c}}{\rho_{c0}}+\Omega_{de0}\frac{\rho_{de}}{\rho_{de0}}
+\Omega_{r0}\frac{\rho_{r}}{\rho_{r0}}+\Omega_{b0}\frac{\rho_{b}}{\rho_{b0}}+\Omega_{k0}\frac{\rho_{k}}{\rho_{k0}},
\end{split}
\end{equation}
where $\Omega_{c0}$, $\Omega_{de0}$, $\Omega_{r0}$, $\Omega_{b0}$ and $\Omega_{k0}$
are the present fractional densities of CDM, DE, radiation, baryon and curvature, respectively.
Since $\Omega_{c0}+\Omega_{de0}+\Omega_{r0}+\Omega_{b0}+\Omega_{k0}=1$,
we do not treat $\Omega_{de0}$ as an independent parameter in this paper.
In addition, $\rho_{r}=\rho_{r0}(1+z)^{4}$, $\rho_{b}=\rho_{b0}(1+z)^{3}$, $\rho_{k}=\rho_{k0}(1+z)^{2}$,
$\Omega_{r0}=\Omega_{m0} / (1+z_{\rm eq})$,
where $\Omega_{m0} = \Omega_{c0}+\Omega_{b0}$
and $z_{\rm eq}=2.5\times 10^4 \Omega_{m0} h^2 (T_{\rm cmb}/2.7\,{\rm K})^{-4}$
(here we take $T_{\rm cmb}=2.7255\,{\rm K}$).

Considering the interaction between dark sectors,
the dynamical evolutions of CDM and DE become
\begin{eqnarray}
\label{eq:CEt1}&& \dot \rho_c+3H\rho_c=Q,\ \ \\
\label{eq:CEt2}&& \dot \rho_{de}+3H(\rho_{de}+p_{de})=-Q,
\end{eqnarray}
where the over dot denotes the derivative with respect to the cosmic time $t$,
$p_{de} = w\rho_{de}$ is the pressure of DE, $w$ is the equation of state of DE,
and $Q$ is the interaction term, which describes the energy transfer rate between CDM and DE.
Notice that $a=\frac{1}{1+z}$ and $H=\frac{\dot a}{a}$, we have $\frac{d}{dt}=-H(1+z)\frac{d}{dz}$.
Then Eqs. (\ref{eq:CEt1}) and (\ref{eq:CEt2}) can be rewritten as
\begin{eqnarray}
\label{eq:CEz6}&&(1+z)\frac{d\rho_{c}}{dz}-3\rho_{c}=-Q/H,\ \ \\
\label{eq:CEz7}&&(1+z)\frac{d\rho_{de}}{dz}-3(1+w)\rho_{de}=Q/H.
\end{eqnarray}
The solutions of these two equations depend on the specific forms of $Q$.

So far, the microscopic origin of interaction between dark sectors is still a big puzzle to us.
To study the issue of interaction,
one has to write down the possible forms of $Q$ by hand.
In this paper we consider the following four cases:
\begin{eqnarray}
    &&Q_{0}=0,\ \ \\
    &&Q_{1}=3\gamma H\rho_{c},\ \ \\
    &&Q_{2}=3\gamma H\rho_{de},\ \ \\
    &&Q_{3}=3\gamma H\frac{\rho_{c}\rho_{de}}{\rho_{c}+\rho_{de}},\ \
\end{eqnarray}
where $\gamma$ is a dimensionless coupling parameter describing the strength of interaction.
Notice that the model with $Q_{0}$ denotes the case without dark sector interaction;
the models with $Q_{1}$ and $Q_{2}$ are very popular,
and both of them have been widely studied in the literature (see, e.g., \cite{IDEPaper1,IDEPaper2,IDEPaper3,IDEPaper4});
the model with $Q_{3}$ is proposed in Ref. \cite{IDEP2},
and it can solve the early-time superhorizon instability and future unphysical CDM density problems at the same time.
For simplicity, hereafter we call them $w$CDM model, I$w$CDM1 model,  I$w$CDM2 model, and I$w$CDM3 model, respectively.

%==================== Q_{0}=0 ====================

For the $w$CDM model, the reduced Hubble parameter $E(z)\equiv H(z)/H_{0}$ can be written as
\begin{equation}\label{E:Q0}
\begin{split}
E(z)=&\big (\Omega_{r0}(1+z)^{4}+(\Omega_{c0}+\Omega_{b0})(1+z)^{3}+\Omega_{k0}(1+z)^{2} +\Omega_{de0}(1+z)^{3(1+w)}\big )^{1/2}.
\end{split}
\end{equation}

%==================== Q_{1}=3\gamma H\rho_{c} ====================

For the I$w$CDM1 model, Eq. (\ref{eq:CEz6}) has a general solution
\begin{equation}\label{rho:c1}
    \rho_{c}=\rho_{c0}(1+z)^{3(1-\gamma)}.
\end{equation}
Substituting Eq. (\ref{rho:c1}) into Eq. (\ref{eq:CEz7}) and using the initial condition $\rho_{de}(z=0)=\rho_{de0}$ ,we get
\begin{equation}\label{rho:de1}
    \rho_{de}=\frac{\gamma\rho_{c0}}{\gamma+w}\big((1+z)^{3(1+w)}-(1+z)^{3(1-\gamma)}\big)+\rho_{de0}(1+z)^{3(1+w)}.
\end{equation}
Then, substituting Eqs. (\ref{rho:c1}) and (\ref{rho:de1}) into Eq. (\ref{E}), we obtain
\begin{equation}\label{E:Q1}
\begin{split}
E(z)=&\Big (\Omega_{r0}(1+z)^{4}+\Omega_{b0}(1+z)^{3}+\Omega_{k0}(1+z)^{2}+\Omega_{de0}(1+z)^{3(1+w)}
+\Omega_{c0}\big (\frac{\gamma}{w+\gamma}(1+z)^{3(1+w)}+\frac{w}{w+\gamma}(1+z)^{3(1-\gamma)}\big )\Big )^{1/2}.
\end{split}
\end{equation}
\\

%==================== Q_{2}=3\gamma H\rho_{de} ====================

For the I$w$CDM2 model, Eq. (\ref{eq:CEz7}) has a general solution
\begin{equation}\label{rho:de2}
    \rho_{de}=\rho_{de0}(1+z)^{3(1+w+\gamma)}.
\end{equation}
Substituting Eq. (\ref{rho:de2}) into Eq. (\ref{eq:CEz6}) and using the initial condition $\rho_{c}(z=0)=\rho_{c0}$, we get
\begin{equation}\label{rho:c2}
    \rho_{c}=\rho_{c0}(1+z)^{3}+\frac{\gamma\rho_{de0}}{w+\gamma}(1+z)^{3}-\frac{\gamma\rho_{de0}}{w+\gamma}(1+z)^{3(1+w+\gamma)}.
\end{equation}
Then, substituting Eqs. (\ref{rho:de2}) and (\ref{rho:c2}) into Eq. (\ref{E}), we have
\begin{equation}\label{E:Q2}
\begin{split}
    E(z)=&\Big (\Omega_{r0}(1+z)^{4}+(\Omega_{c0}+\Omega_{b0})(1+z)^{3}+\Omega_{k0}(1+z)^{2}
    +\Omega_{de0}\big (\frac{\gamma}{w+\gamma}(1+z)^{3} + \frac{w}{w+\gamma}(1+z)^{3(1+w+\gamma)} \big )\Big )^{1/2}.
\end{split}
\end{equation}
\\

%========== Q_{3}=3\gamma H\frac{\rho_{c}\rho_{de}}{\rho_{c}+\rho_{de}} ==========

For the I$w$CDM3 model,
the energy densities of CDM and DE satisfy
\begin{equation}\label{rho:c3}
\begin{split}
\rho_{c}=&\rho_{c0}(1+z)^{3}\big(\frac{\rho_{c0}}{\rho_{c0}+\rho_{de0}} +\frac{\rho_{de0}}{\rho_{c0}+\rho_{de0}}(1+z)^{3(w+\gamma)}\big)^{-\frac{\gamma}{w+\gamma}},
\end{split}
\end{equation}
\begin{equation}\label{rho:de3}
\begin{split}
\rho_{de}=&\rho_{de0}(1+z)^{3(1+w+\gamma)}\big(\frac{\rho_{c0}}{\rho_{c0}+\rho_{de0}} +\frac{\rho_{de0}}{\rho_{c0}+\rho_{de0}}(1+z)^{3(w+\gamma)}\big)^{-\frac{\gamma}{w+\gamma}}.
\end{split}
\end{equation}
Substituting Eqs. (\ref{rho:c3}) and (\ref{rho:de3}) into Eq. (\ref{E}), we get
\begin{equation}\label{E:Q3}
\begin{split}
     E(z)=&\big(\Omega_{c0}C(z)(1+z)^{3}+\Omega_{de0}C(z)(1+z)^{3(1+w+\gamma)}
     +\Omega_{r0}(1+z)^{4}+\Omega_{b0}(1+z)^{3}+\Omega_{k0}(1+z)^{2}\big)^{1/2},
\end{split}
\end{equation}
where
\begin{equation}\label{Cz}
C(z)=\big(\frac{\Omega_{c0}}{\Omega_{c0}+\Omega_{de0}}+\frac{\Omega_{de0}}{\Omega_{c0}+\Omega_{de0}}(1+z)^{3(w+\gamma)}\big)^{-\frac{\gamma}{w+\gamma}}.
\end{equation}

Note that in Eqs. (\ref{E:Q0}), (\ref{E:Q1}), (\ref{E:Q2}), (\ref{E:Q3}), and (\ref{Cz}),
$\Omega_{de0}$ is not an independent parameter, which is given by
$\Omega_{de0}=1-\Omega_{c0}-\Omega_{b0}-\Omega_{r0}-\Omega_{k0}$.

\subsection{Observational data}

In this subsection, we introduce how to include the SNLS3 data into the $\chi^2$ analysis.

For the SNLS3 sample, the observable is $m_B$, which is the rest-frame peak B-band magnitude of the SN.
By considering three functional forms (linear case, quadratic case, and step function case),
Wang \& Wang \cite{WangWang} showed that the evolutions of $\alpha$ and $\beta$
are insensitive to functional form of $\alpha$ and $\beta$ assumed.
So in this paper,
we just adopt a constant $\alpha$ and a linear $\beta(z) = \beta_{0} + \beta_{1} z$.
Then, the predicted magnitude of an SN becomes
\be
m_{\rm mod}=5 \log_{10}{\cal D}_L(z)
- \alpha (s-1) +\beta(z) {\cal C} + {\cal M},
\ee
where $s$ and ${\cal C}$ are the stretch measure and the color measure for the SN light curve.
Here ${\cal M}$ is a parameter representing some combination of SN absolute magnitude $M$ and Hubble constant $H_0$.
It must be emphasized that,
to include host-galaxy information in the cosmological fits,
Conley et al. \cite{Conley11} split the SNLS3 sample based on host-galaxy stellar mass at $10^{10} M_{\odot}$,
and made ${\cal M}$ to be different for the two samples.
Therefore, unlike other SN samples, there are two values of ${\cal M}$, ${\cal M}_1$ and ${\cal M}_2$, for the SNLS3 data
Moreover, Conley et al. removed ${\cal M}_1$ and ${\cal M}_2$ from cosmology-fits by analytically marginalizing over them
(for more details, see the appendix C of \cite{Conley11},
as well as the the public code which is available at https://tspace.library.utoronto.ca/handle/1807/24512).
In this paper, we just follow the recipe of Ref. \cite{Conley11}.
The luminosity distance ${\cal D}_L(z)$ is defined as
\be
{\cal D}_L(z)\equiv H_0 (1+z_{\rm hel}) r(z),
\ee
where $z$ and $z_{\rm hel}$ are the CMB restframe and heliocentric redshifts of SN.
In addition, the comoving distance $r(z)$ is given by
\be
\label{eq:r(z)}
 r(z)=H_0^{-1}\, |\Omega_{k0}|^{-1/2} {\rm sinn}\big (|\Omega_{k0}|^{1/2}\, \Gamma(z)\big ),
\ee
where $\Gamma(z)=\int_0^z\frac{dz'}{E(z')}$,
and ${\rm sinn}(x)=\sin(x)$, $x$, $\sinh(x)$ for $\Omega_{k0}<0$, $\Omega_{k0}=0$, and $\Omega_{k0}>0$ respectively.

For a set of $N$ SNe with correlated errors, the $\chi^2$ function is
\be
\label{eq:chi2_SN}
\chi^2_{SN}=\Delta \mbox{\bf m}^T \cdot \mbox{\bf C}^{-1} \cdot \Delta\mbox{\bf m},
\ee
where $\Delta m \equiv m_B-m_{\rm mod}$ is a vector with $N$ components,
and $\mbox{\bf C}$ is the $N\times N$ covariance matrix of the SN, given by
\be
\mbox{\bf C}=\mbox{\bf D}_{\rm stat}+\mbox{\bf C}_{\rm stat}+\mbox{\bf C}_{\rm sys}.
\ee
$\mbox{\bf D}_{\rm stat}$ is the diagonal part of the statistical uncertainty, given by \cite{Conley11}
\ba
\mbox{\bf D}_{{\rm stat},ii}&=&\sigma^2_{m_B,i}+\sigma^2_{\rm int}
+ \sigma^2_{\rm lensing}+ \sigma^2_{{\rm host}\,{\rm correction}} + \left[\frac{5(1+z_i)}{z_i(1+z_i/2)\ln 10}\right]^2 \sigma^2_{z,i} \nonumber\\
&& +\alpha^2 \sigma^2_{s,i}+\beta(z_i)^2 \sigma^2_{{\cal C},i} + 2 \alpha C_{m_B s,i} - 2 \beta(z_i) C_{m_B {\cal C},i} -2\alpha \beta(z_i) C_{s {\cal C},i},
\ea
where $C_{m_B s,i}$, $C_{m_B {\cal C},i}$, and $C_{s {\cal C},i}$
are the covariances between $m_B$, $s$, and ${\cal C}$ for the $i$-th SN,
$\beta_i=\beta(z_i)$ are the values of $\beta$ for the $i$-th SN.
Notice that $\sigma^2_{z,i}$ includes a peculiar velocity residual of 0.0005
(i.e., 150$\,$km/s) added in quadrature.
Following the Ref. \cite{Conley11}, we fix the intrinsic scatter $\sigma_{int}$ to ensure that $\chi^2/dof=1$.
Varying $\sigma_{int}$ could have a significant impact on parameter estimation, see \cite{Kim2011} for details.

We define $\mbox{\bf V} \equiv \mbox{\bf C}_{\rm stat} + \mbox{\bf C}_{\rm sys}$,
where $\mbox{\bf C}_{\rm stat}$ and $\mbox{\bf C}_{\rm sys}$
are the statistical and systematic covariance matrices, respectively.
After treating $\beta$ as a function of $z$,
$\mbox{\bf V}$ is given in the form,
\ba
\mbox{\bf V}_{ij}&=&V_{0,ij}+\alpha^2 V_{a,ij} + \beta_i\beta_j V_{b,ij} +\alpha V_{0a,ij} +\alpha V_{0a,ji}
-\beta_j V_{0b,ij} -\beta_i V_{0b,ji} -\alpha \beta_j V_{ab,ij} - \alpha \beta_i V_{ab,ji}.
\ea
It must be stressed that, while $V_0$, $V_{a}$, $V_{b}$, and $V_{0a}$
are the same as the ``normal'' covariance matrices
given by the SNLS data archive, $V_{0b}$, and $V_{ab}$ are {\it not} the same as the ones given there.
This is because the original matrices of SNLS3 are produced by assuming $\beta$ is constant.
We have used the $V_{0b}$ and $V_{ab}$ matrices for the ``Combined'' set
that are applicable when varying $\beta(z)$ (A.~Conley, private communication, 2013).

To improve the cosmological constraints,
we also use some other cosmological observations,
including the Planck distance prior data \cite{WangWangCMB},
the galaxy clustering (GC) data extracted from SDSS DR7 \cite{ChuangWang12} and BOSS \cite{Chuang13},
as well as the direct measurement of Hubble constant $H_0=73.8\pm 2.4 {\rm km/s/Mpc}$ from the HST observations \cite{Riess11}.
For the details of including Planck and GC data into the $\chi^2$ analysis, see Ref. \cite{WangNew}.
Now the total $\chi^2$ function is
\be
\chi^2=\chi^2_{SN}+\chi^2_{CMB}+\chi^2_{GC}+\chi^2_{H_0}.
\ee
In addition, assuming the measurement errors are Gaussian, the likelihood function satisfies
\be
{\cal{L }} \propto e^{-\chi^{2}/2}, \ \ \  \mathrm{Likelihood} \equiv {\cal{L }}/{\cal{L }}_{max} = {\cal{L }}/{\cal{L }}(\chi^{2}_{min}).
\ee

It should be mentioned that, in this paper we just use the purely geometric measurements,
and do not consider the cosmological perturbations in the DE models.
As analyzed in detail in Ref. \cite{IDEP1},
adopting a new framework for calculating the perturbations,
the cosmological perturbations will always be stable in all IDE models
(For a related discussion concerning the stability, see Ref. \cite{IDEP2}).
Therefore, the use of the Planck distance prior is sufficient for our purpose.

Finally, we perform an MCMC likelihood analysis \cite{COSMOMC}
to obtain ${\cal O}$($10^6$) samples for each model considered in this paper.

\section{Results}

\subsection{Evolution of $\beta$}
\label{sec:SNeonly}

In this subsection, we explore the evolution of $\beta$ in the frame of IDE.

In Table \ref{table1}, we list the fitting results for various constant $\beta$ and linear $\beta(z)$ cases,
where the SNe+CMB+GC+$H_0$ data are used.
An obvious feature of this table is that varying $\beta$ can significantly improve the fitting results:
for all the models, adding a parameter of $\beta$ can reduce the best-fit values of $\chi^2$ by $\sim$ 34.
Based on the Wilk's theorem, 34 units of $\chi^2$ is equivalent to a Gaussian fluctuation of 5.8$\sigma$.
This means that the result of $\beta_1=0$ is ruled out at 5.8$\sigma$ confidence level (CL) .
As shown in Refs. \cite{WangNew} and \cite{WangNew2},
for the cases of various DE models (such as $\Lambda$CDM, $w$CDM, CPL, and HDE model) without interaction,
$\beta$ deviates from a constant at 6$\sigma$ CL.
Therefore, we further confirm the redshift-evolution of $\beta$ for the SNLS3 data.

\begin{table*}\tiny
\caption{Fitting results for various constant $\beta$ and linear $\beta(z)$ cases,
where the SNe+CMB+GC+$H_0$ data are used.}
\label{table1}
\begin{tabular}{cccccccccccc}
\hline\hline &\multicolumn{2}{c}{$w$CDM}&&\multicolumn{2}{c}{I$w$CDM1}&&\multicolumn{2}{c}{I$w$CDM2}&&\multicolumn{2}{c}{I$w$CDM3} \\
           \cline{2-3}\cline{5-6}\cline{8-9}\cline{11-12}
Parameters  & Const $\beta$ & Linear $\beta(z)$ && Const $\beta$ & Linear $\beta(z)$ && Const $\beta$ & Linear $\beta(z)$ && Const $\beta$ & Linear $\beta(z)$\\ \hline
$\alpha$           & $1.444^{+0.079}_{-0.115}$
                   & $1.423^{+0.087}_{-0.095}$&
                   & $1.424^{+0.104}_{-0.094}$
                   & $1.398^{+0.110}_{-0.066}$&
                   & $1.427^{+0.096}_{-0.097}$
                   & $1.421^{+0.084}_{-0.101}$&
                   & $1.445^{+0.082}_{-0.115}$
                   & $1.393^{+0.121}_{-0.068}$\\

$\beta_0$          & $3.251^{+0.113}_{-0.098}$
                   & $1.518^{+0.326}_{-0.378}$&
                   & $3.272^{+0.087}_{-0.116}$
                   & $1.438^{+0.367}_{-0.372}$&
                   & $3.275^{+0.084}_{-0.112}$
                   & $1.474^{+0.308}_{-0.369}$&
                   & $3.248^{+0.110}_{-0.084}$
                   & $1.505^{+0.292}_{-0.402}$\\

$\beta_1$          &
                   & $4.926^{+1.011}_{-0.869}$&
                   &
                   & $5.102^{+0.988}_{-0.924}$&
                   &
                   & $4.970^{+1.015}_{-0.819}$&
                   &
                   & $4.886^{+1.191}_{-0.747}$\\

$\Omega_{c0}$      & $0.224^{+0.010}_{-0.010}$
                   & $0.231^{+0.011}_{-0.009}$&
                   & $0.232^{+0.012}_{-0.015}$
                   & $0.244^{+0.016}_{-0.013}$&
                   & $0.226^{+0.012}_{-0.013}$
                   & $0.238^{+0.020}_{-0.012}$&
                   & $0.225^{+0.014}_{-0.011}$
                   & $0.244^{+0.013}_{-0.016}$\\

$\Omega_{b0}$      & $0.042^{+0.002}_{-0.002}$
                   & $0.044^{+0.002}_{-0.002}$&
                   & $0.041^{+0.003}_{-0.002}$
                   & $0.040^{+0.003}_{-0.002}$&
                   & $0.041^{+0.003}_{-0.002}$
                   & $0.041^{+0.003}_{-0.002}$&
                   & $0.042^{+0.002}_{-0.002}$
                   & $0.042^{+0.002}_{-0.003}$\\

$\Omega_{k0}$      & $0.00046^{+0.004}_{-0.003}$
                   & $0.0032^{+0.0038}_{-0.0041}$&
                   & $0.0039^{+0.0044}_{-0.0061}$
                   & $0.0095^{+0.0050}_{-0.0059}$&
                   & $0.0061^{+0.0142}_{-0.0162}$
                   & $0.0192^{+0.0180}_{-0.0165}$&
                   & $0.0046^{+0.0160}_{-0.0131}$
                   & $0.0194^{+0.0194}_{-0.0159}$\\

$\gamma$           &
                   & &
                   & $-0.0028^{+0.0043}_{-0.0031}$
                   & $-0.0053^{+0.0035}_{-0.0026}$&
                   & $-0.0105^{+0.0310}_{-0.0295}$
                   & $-0.0322^{+0.0300}_{-0.0396}$&
                   & $-0.0198^{+0.0613}_{-0.0752}$
                   & $-0.0732^{+0.0684}_{-0.0823}$\\

$w$                & $-1.118^{+0.065}_{-0.071}$
                   & $-1.042^{+0.068}_{-0.072}$&
                   & $-1.105^{+0.075}_{-0.069}$
                   & $-1.016^{+0.075}_{-0.063}$&
                   & $-1.124^{+0.070}_{-0.062}$
                   & $-1.052^{+0.070}_{-0.068}$&
                   & $-1.116^{+0.059}_{-0.072}$
                   & $-1.038^{+0.068}_{-0.080}$\\

$h$                & $0.725^{+0.014}_{-0.014}$
                   & $0.716^{+0.014}_{-0.015}$&
                   & $0.739^{+0.019}_{-0.023}$
                   & $0.743^{+0.016}_{-0.024}$&
                   & $0.734^{+0.018}_{-0.025}$
                   & $0.735^{+0.017}_{-0.024}$&
                   & $0.732^{+0.022}_{-0.021}$
                   & $0.729^{+0.027}_{-0.018}$\\

\hline $\chi^{2}_{min}$  & 422.696  & 388.508 && 422.376 & 387.128 && 422.674 & 387.814 && 422.642 & 387.716 \\
\hline
\end{tabular}
\end{table*}

\begin{figure}
\psfig{file=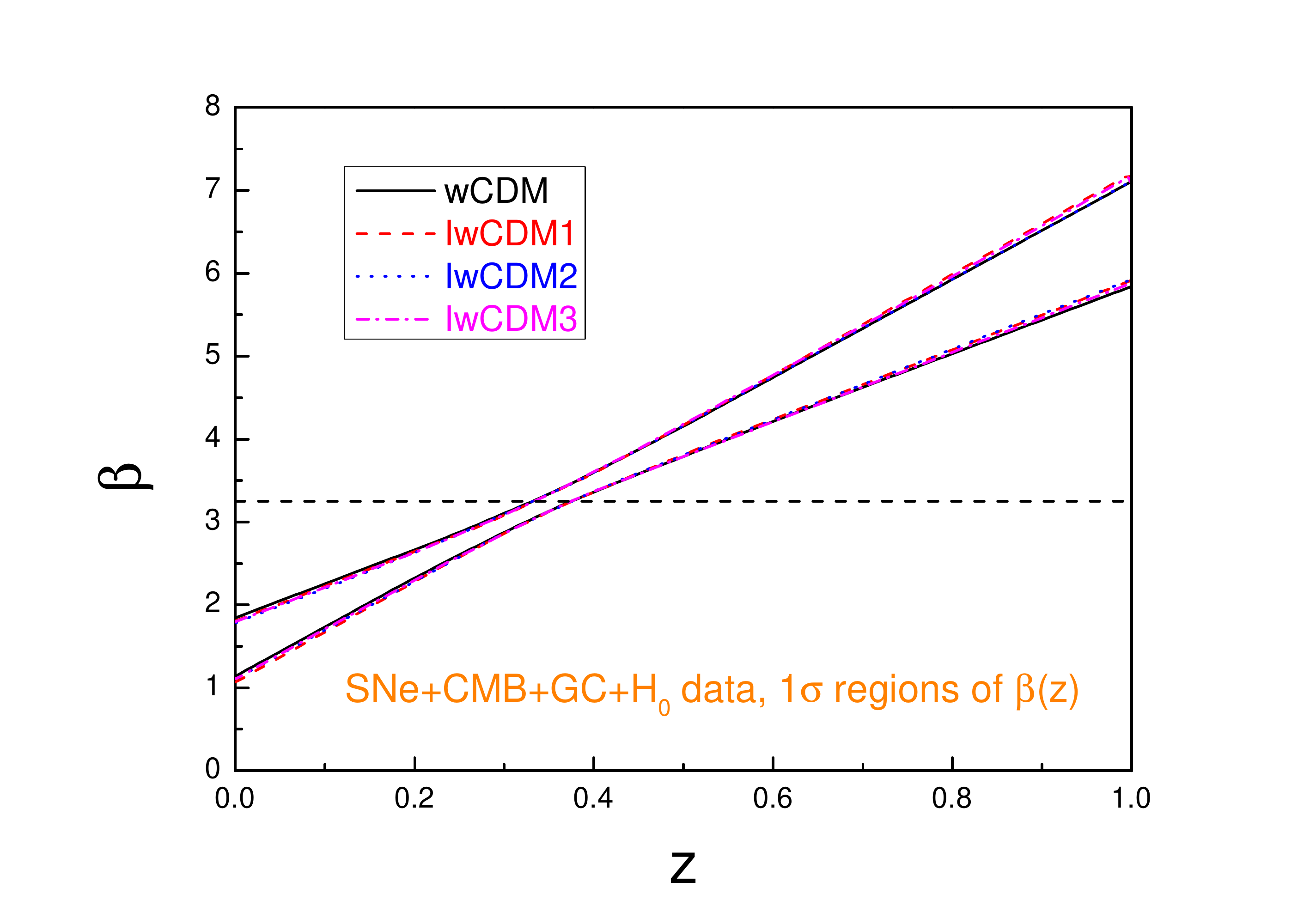,width=3.5in}\\
\caption{\label{fig1}\footnotesize%
1$\sigma$ confidence constraints for the evolution of $\beta(z)$,
given by the SNe+CMB+GC+$H_0$ data, for the $w$CDM model, the I$w$CDM1 model, the I$w$CDM2 model, and the I$w$CDM3 model.
The solid black lines denote the $w$CDM model, the dashed red lines denote the I$w$CDM1 model,
the dotted blue lines denote the I$w$CDM2 model, and the dashed-dotted pink lines denote the I$w$CDM3 model.
To make a comparison, for the $w$CDM model, the best-fit result of constant $\beta$ case is also plotted, shown as the horizontal dashed black line.}
\end{figure}

In Fig. \ref{fig1}, using the SNe+CMB+GC+$H_0$ data,
we plot the 1$\sigma$ confidence constraints of $\beta(z)$,
for the $w$CDM model, the I$w$CDM1 model, the I$w$CDM2 model, and the I$w$CDM3 model.
For comparison, we also plot the best-fit result of constant $\beta$ case for the $w$CDM model.
From this figure one can see that,
the 1$\sigma$ regions of $\beta(z)$ of all these models are almost overlapping.
This shows that the evolution of $\beta$ is independent of the interacting dark energy models.
In addition, for all the models, $\beta(z)$ rapidly increases with $z$.
This result is consistent with the results of Refs. \cite{WangNew} and \cite{WangNew2},
showing that the evolution of $\beta$ is insensitive to
dark energy models including those with
interaction between dark sectors.

It should be pointed out that the evolutionary behaviors of $\beta(z)$ depends on the SN samples used.
In \cite{Mohlabeng}, Mohlabeng and Ralston found that, for the Union2.1 SN data, $\beta(z)$ decreases with $z$.
This is similar to the case of Pan-STARRS1 SN data \cite{Scolnic2}.

It is interesting to study how different segments of the SNLS3 dataset give rise to different behavior of $\beta$.
To do this, we perform the following test:
(1) Per \cite{Conley11}, we evenly divide the redshift range $[0, 1]$ into 9 bins and assume that both $\alpha$ and $\beta$ are constant within each bin.
(2) For each redshift bin, we make a small covariance matrix corresponding to only SNe in that bin.
(3) Since we have already proved that the evolution of $\beta$ is insensitive to dark energy models,
per \cite{Conley11}, we just adopt a fixed cosmological background (a flat $\Lambda$CDM model with $\Omega_{m0}$ = 0.26) to do this test.
(4) We fit $\alpha$ and $\beta$ separately for the 9 redshift bins.
Based on the best-fit analysis, it is found that $\beta$ is relatively flat till the 7th bin, and then it rapidly increases along with redshift $z$.
In other words, the rapid increase of $\beta(z)$ is mainly due to the contributions from high-redshift ($z>0.7$) SN samples of SNLS3 dataset.
It should be mentioned that, to keep this paper focus on its main purpose,
here we just briefly present the conclusion, instead of describing all the detailed results of the test.
To understand why high-redshift SNLS3 samples will yield this kind of evolutionary behavior of $\beta$,
some numerical simulation studies may be needed.
We will study this issue in future works.

\subsection{Effects of time-varying $\beta$}
\label{sec:Combindata}

In this subsection, we discuss the effects of varying $\beta$ on parameter estimation of IDE models.

\begin{figure}
\includegraphics[width=0.35\textwidth]{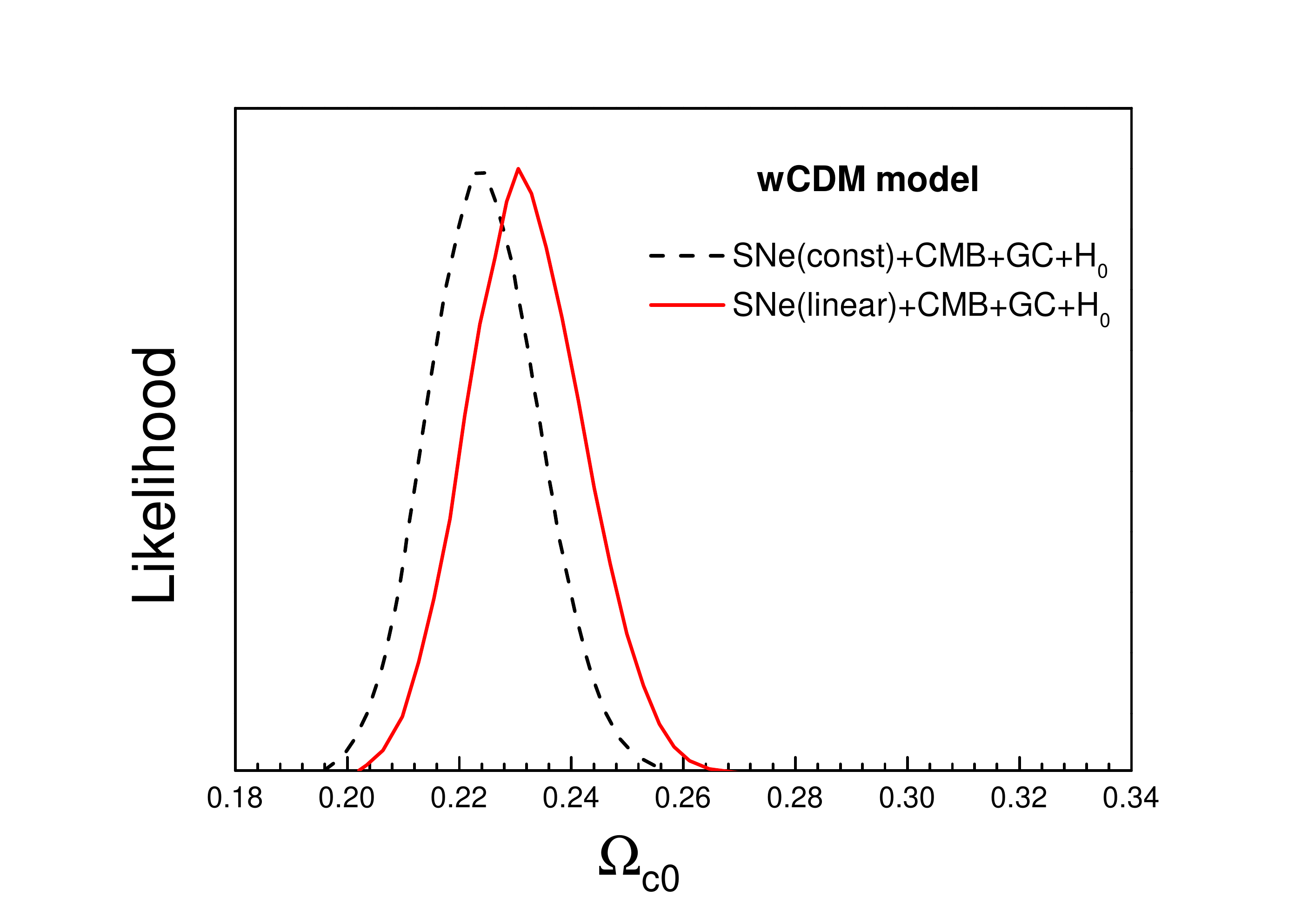}\hskip.4cm
\includegraphics[width=0.35\textwidth]{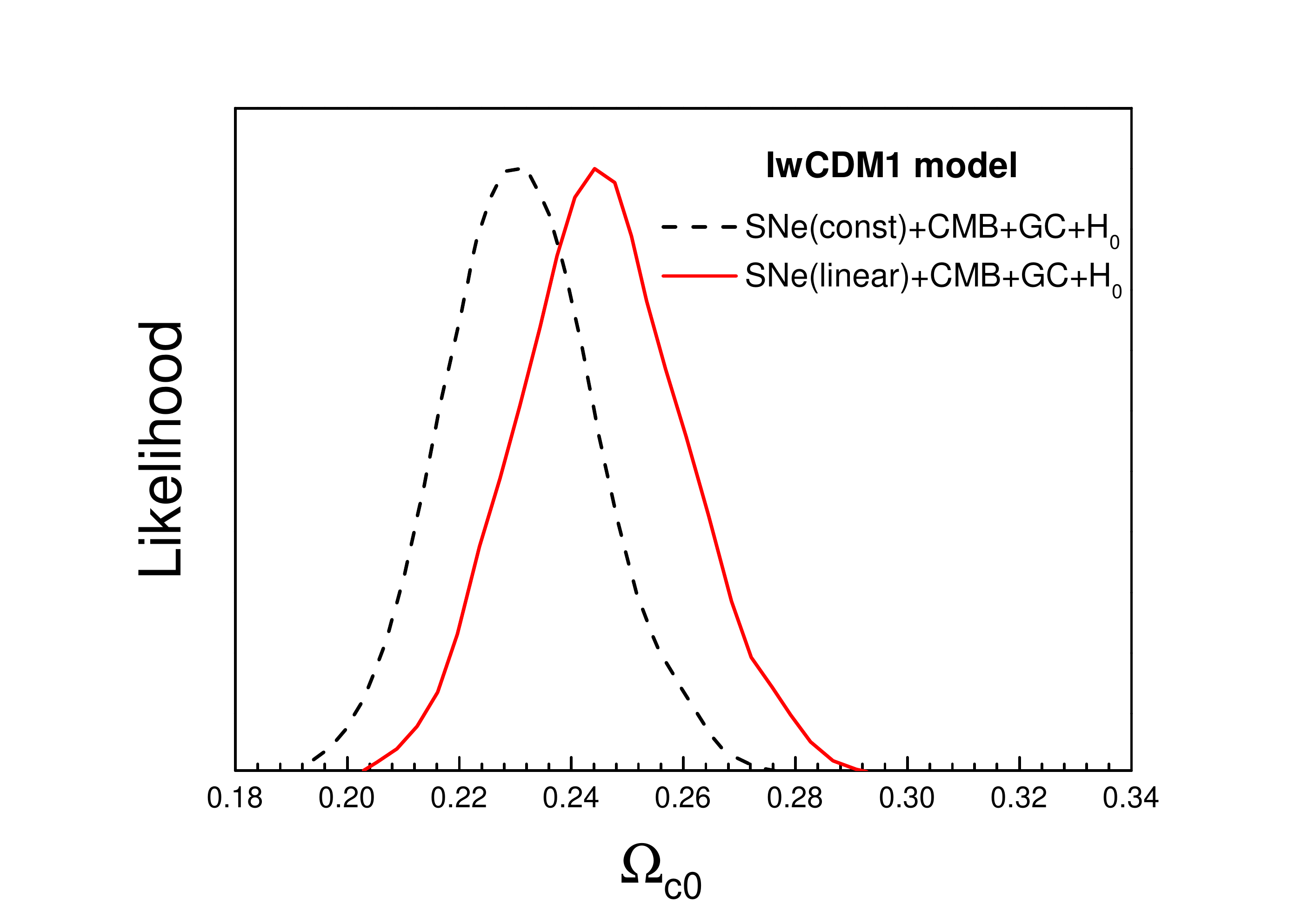}\vskip.4cm
\includegraphics[width=0.35\textwidth]{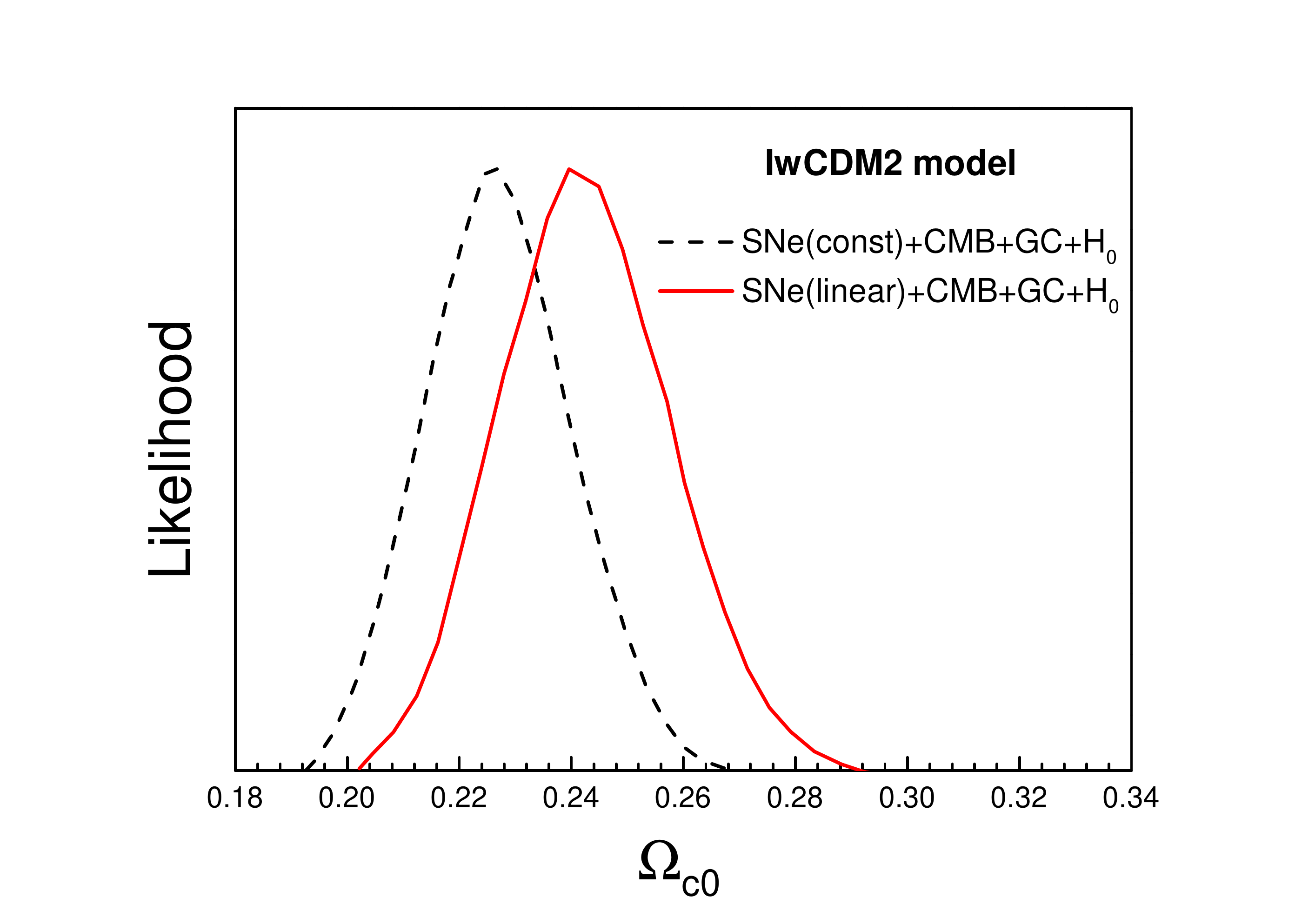}\hskip.4cm
\includegraphics[width=0.35\textwidth]{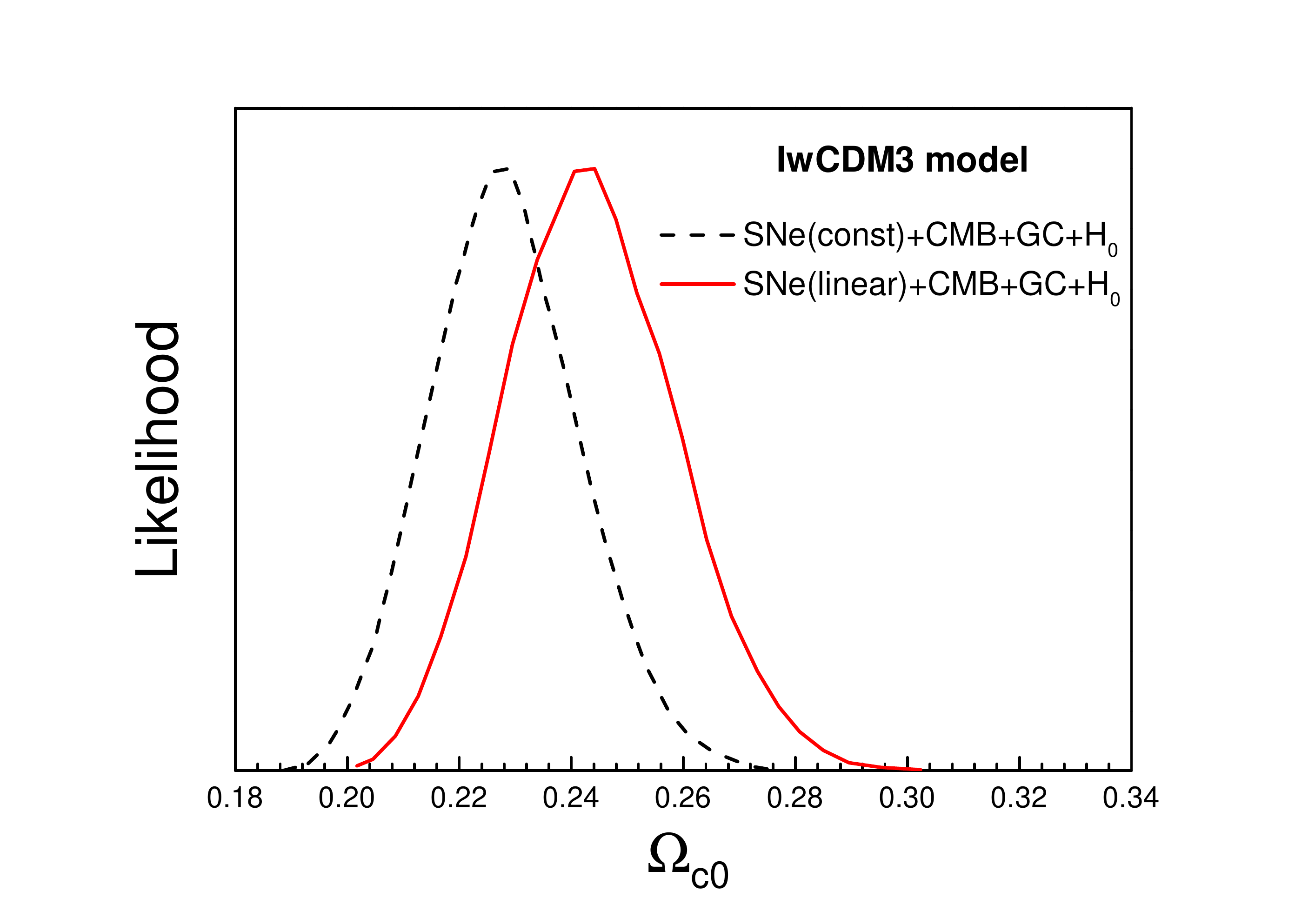}
\caption{\label{fig2}\footnotesize%
The 1D marginalized probability distributions of $\Omega_{c0}$, given by the SNe+CMB+GC+$H_0$ data,
for the $w$CDM model and the three IDE models.
Both the results of constant $\beta$ (dashed black lines) and linear $\beta(z)$ (solid red lines) cases are presented.}
\end{figure}

In Fig. \ref{fig2}, using SNe+CMB+GC+$H_0$ data,
we plot the 1D marginalized probability distributions of $\Omega_{c0}$,
for all the models considered in this paper.
We find that varying $\beta$ yields a larger $\Omega_{c0}$:
for the constant $\beta$ case, the best-fit results are $\Omega_{c0}=0.224$, 0.232, 0.226, and 0.225,
for the $w$CDM, the I$w$CDM1, the I$w$CDM2, and the I$w$CDM3 model, respectively;
while for the linear $\beta(z)$ case, the best-fit results are $\Omega_{c0}=0.231$, 0.244, 0.238, and 0.244,
for the $w$CDM, the I$w$CDM1, the I$w$CDM2, and the I$w$CDM3 model, respectively.
In addition, as shown in Refs. \cite{WangNew} and \cite{WangNew2}, for various DE models without interaction term,
a time-varying $\beta$ also yields a larger fractional matter density $\Omega_{m0} \equiv \Omega_{c0}+\Omega_{b0}$.
Therefore, we can conclude that the effects of varying $\beta$ on the present fractional matter density
are insensitive to the interaction between dark sectors.

\begin{figure}
\includegraphics[width=0.35\textwidth]{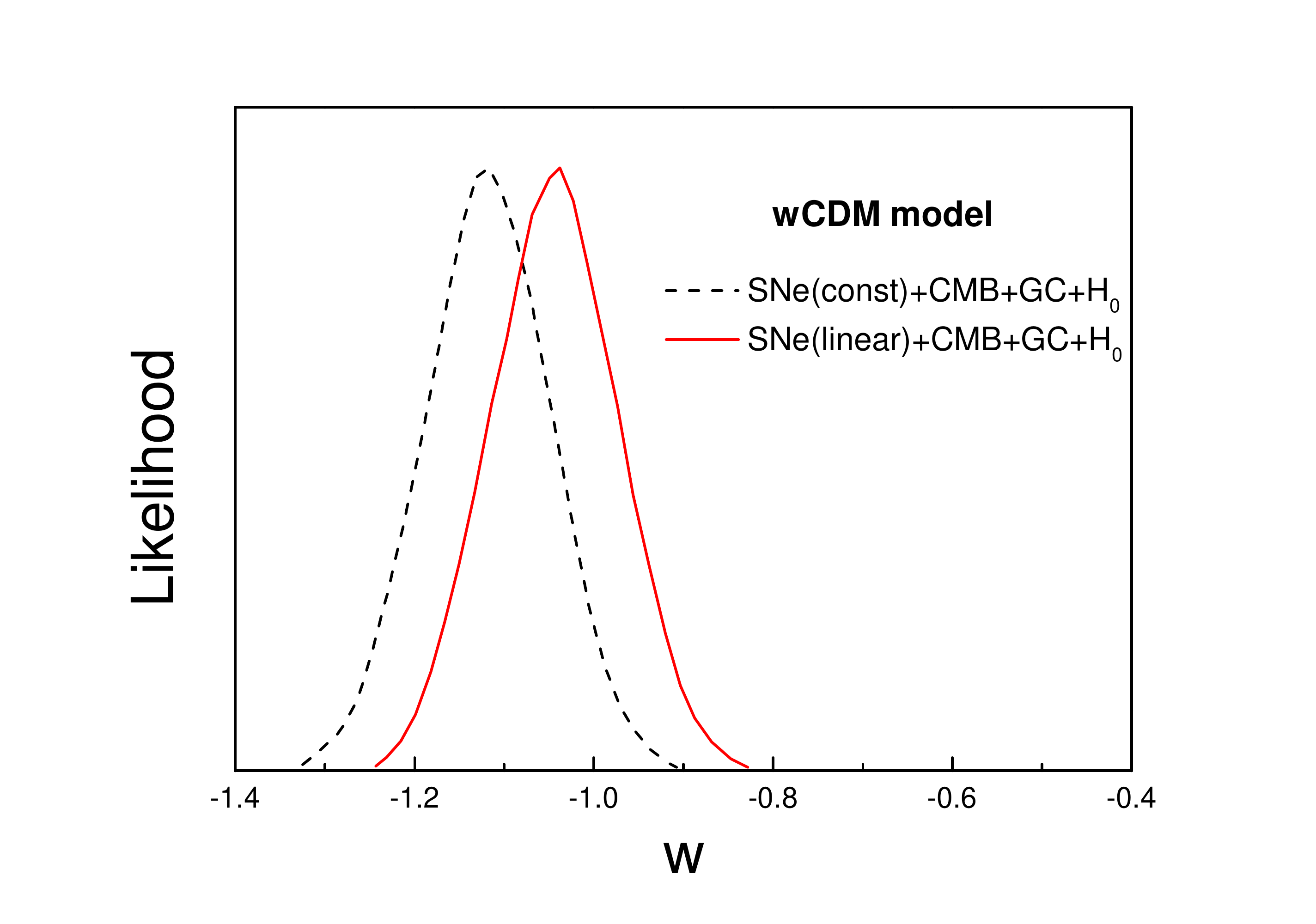}\hskip.4cm
\includegraphics[width=0.35\textwidth]{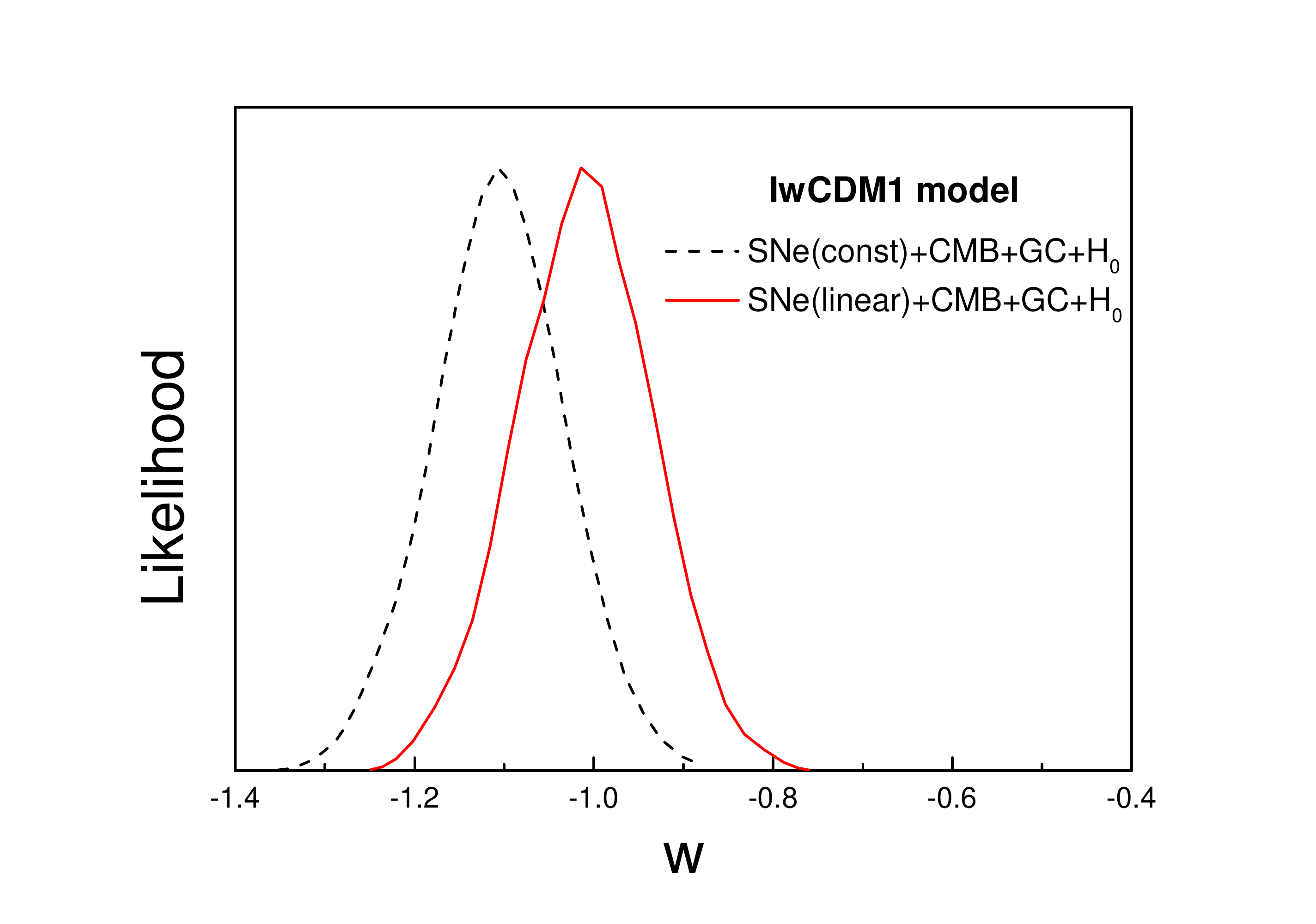}\vskip.4cm
\includegraphics[width=0.35\textwidth]{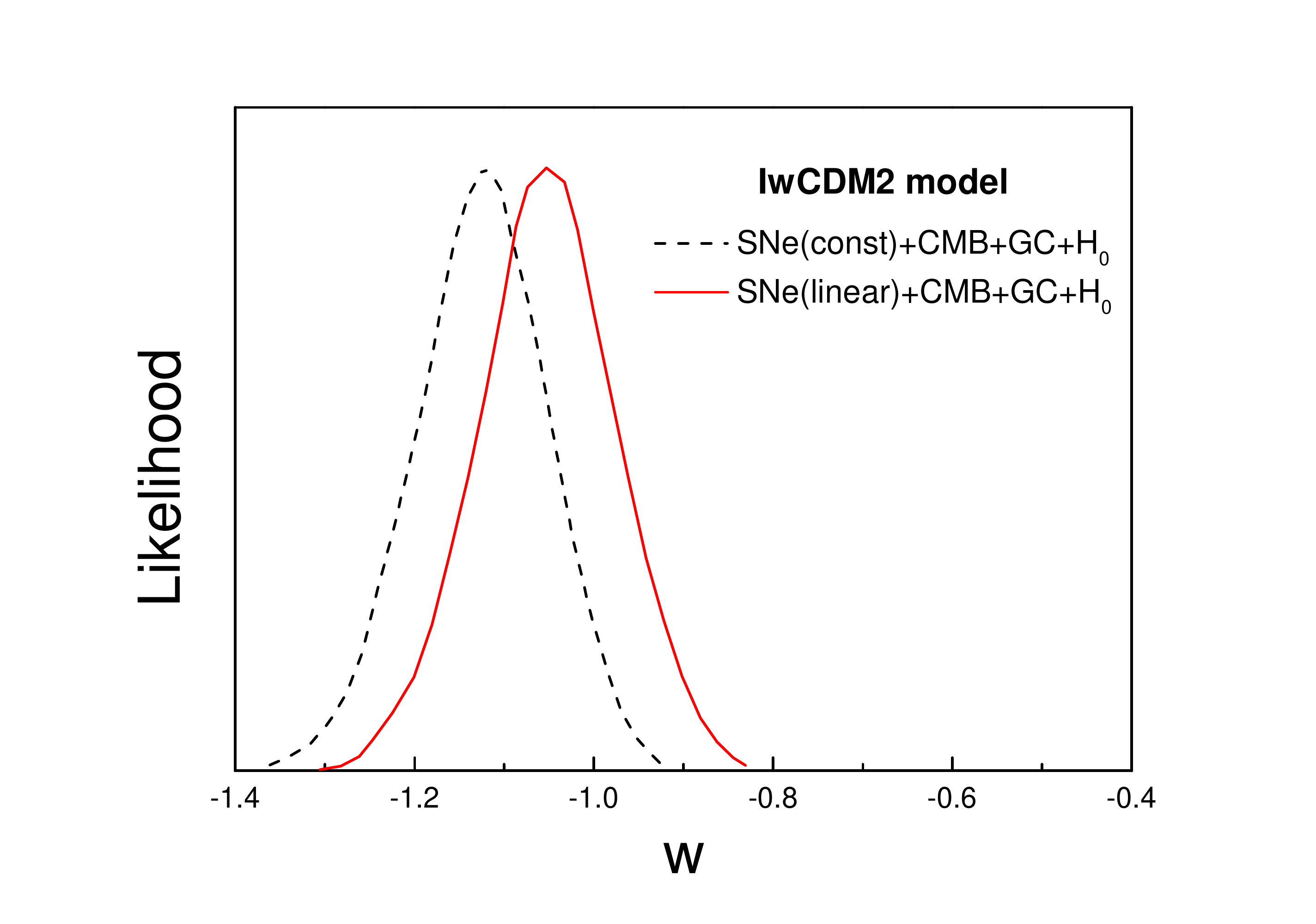}\hskip.4cm
\includegraphics[width=0.35\textwidth]{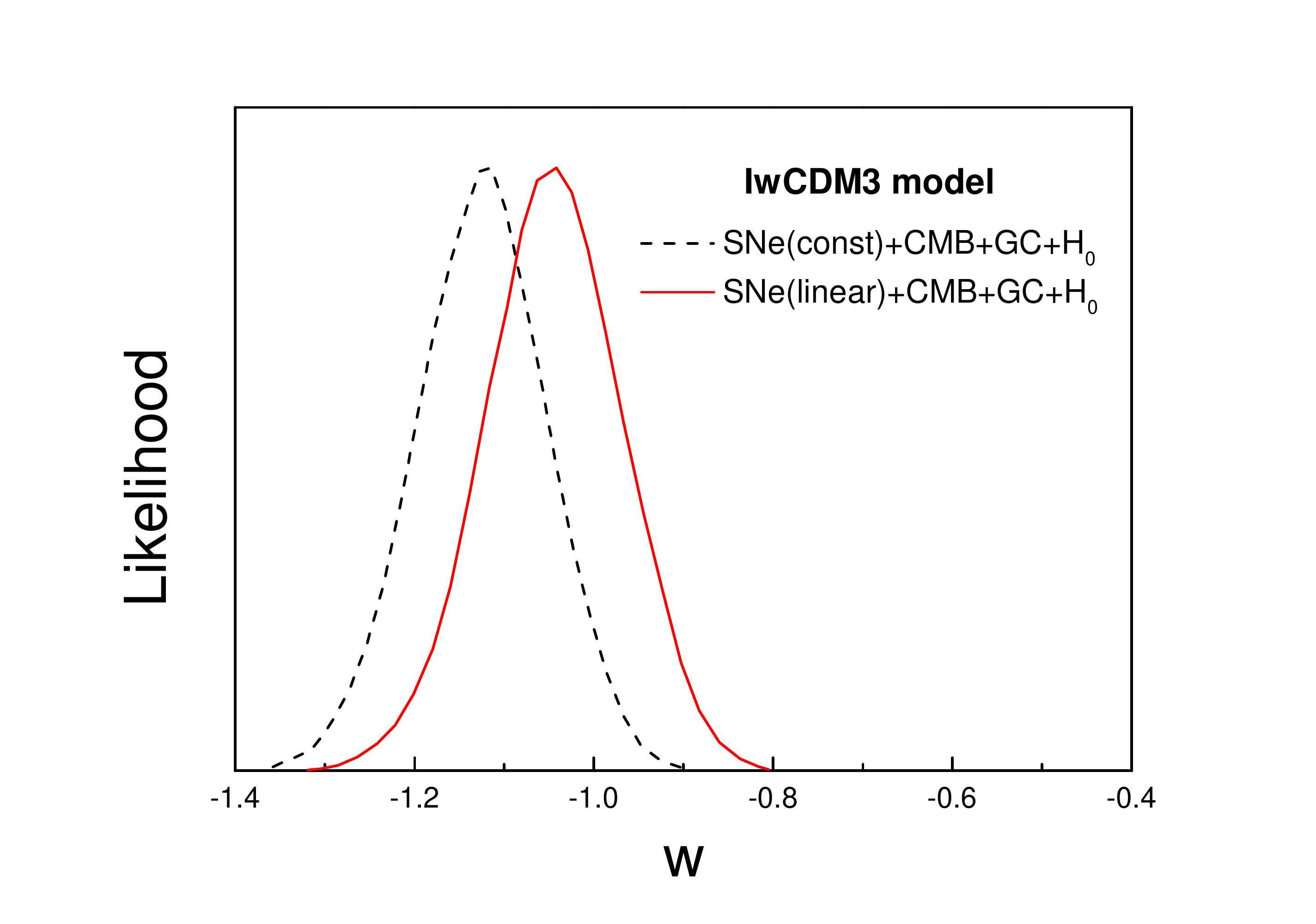}
\caption{\label{fig3}\footnotesize%
The 1D marginalized probability distributions of $w$, given by the SNe+CMB+GC+$H_0$ data,
for the $w$CDM model and the three IDE models.
Both the results of constant $\beta$ (dashed black lines) and linear $\beta(z)$ (solid red lines) cases are presented.}
\end{figure}

For all the models considered in this paper,
the 1D marginalized probability distributions of $w$ are plotted in Fig. \ref{fig3}.
It is found that varying $\beta$ yields a larger $w$:
for the constant $\beta$ case,
$w=-1.118^{+0.065}_{-0.071}, -1.105^{+0.075}_{-0.069}, -1.124^{+0.070}_{-0.062}$, and $-1.116^{+0.059}_{-0.072}$,
for the $w$CDM model, the I$w$CDM1 model, the I$w$CDM2 model, and the I$w$CDM3 model, respectively;
while for the linear $\beta(z)$ case,
$w=-1.042^{+0.068}_{-0.072}, -1.016^{+0.075}_{-0.063}, -1.052^{+0.070}_{-0.068}$, and $-1.038^{+0.068}_{-0.080}$,
for the $w$CDM model, the I$w$CDM1 model, the I$w$CDM2 model, and the I$w$CDM3 model, respectively.
In other words, $w<-1$ is preferred at more than 1$\sigma$ CL for the constant $\beta$ case,
while $w$ is consistent with $-1$ at 1$\sigma$ CL for the linear $\beta(z)$ case.
This means that, compared to the constant $\beta$ case,
the results from varying $\beta$ case are in better agreement with a cosmological constant.
This conclusion is consistent with the noninteracting cases \cite{WangNew,WangNew2},
showing that the effects of varying $\beta$ on $w$ are insensitive to the interaction between dark sectors.

\begin{figure}
\includegraphics[width=0.35\textwidth]{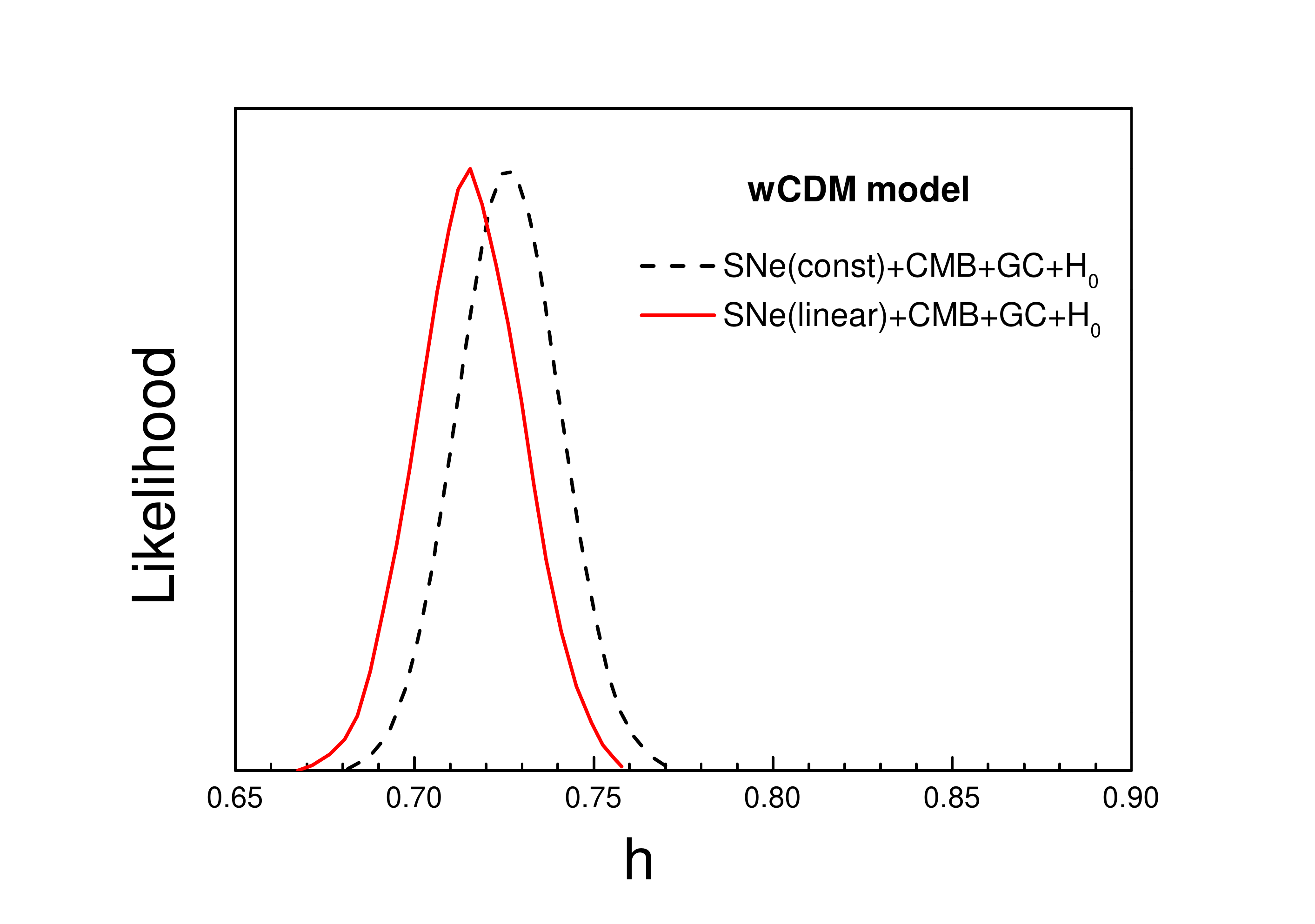}\hskip.4cm
\includegraphics[width=0.35\textwidth]{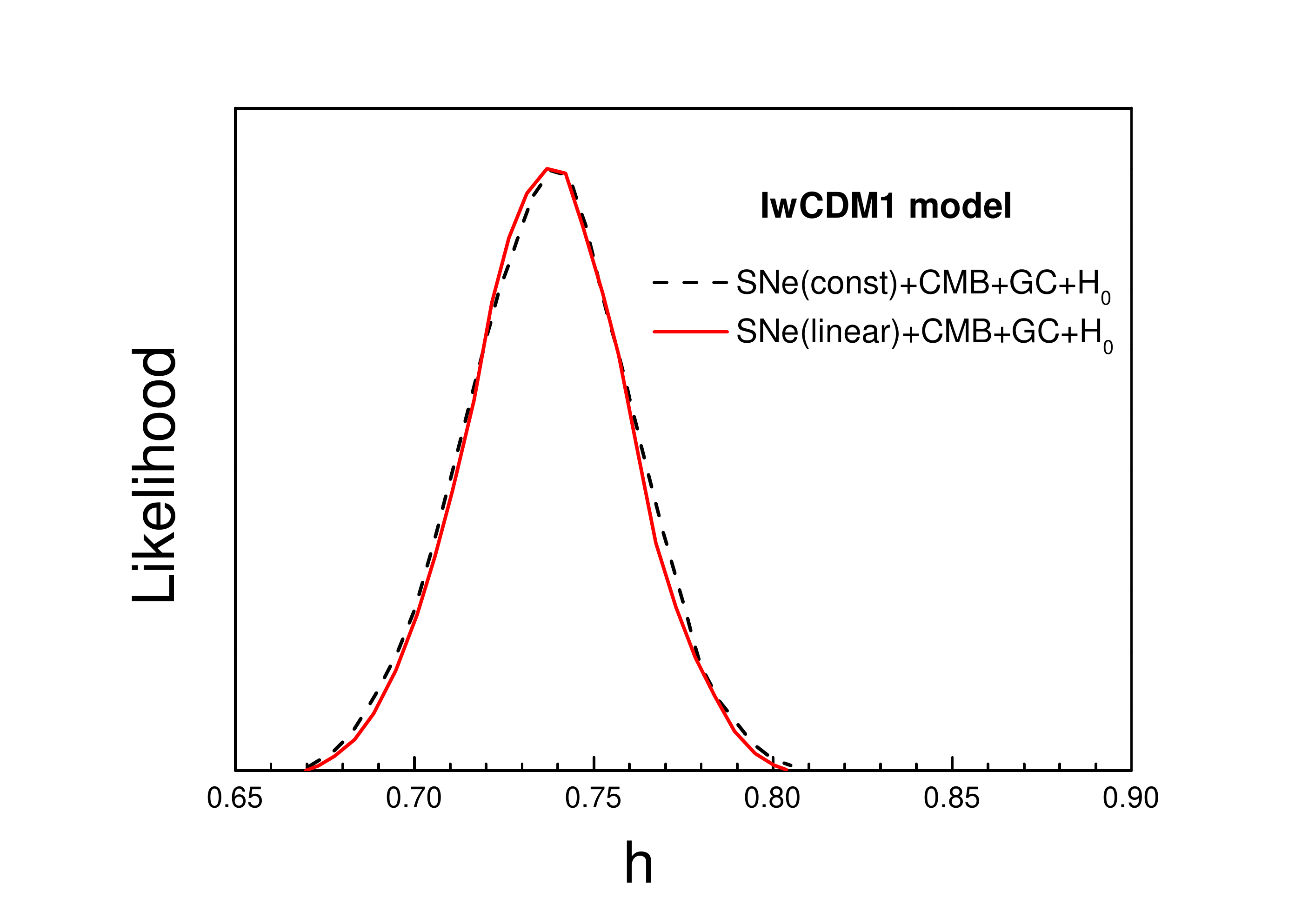}\vskip.4cm
\includegraphics[width=0.35\textwidth]{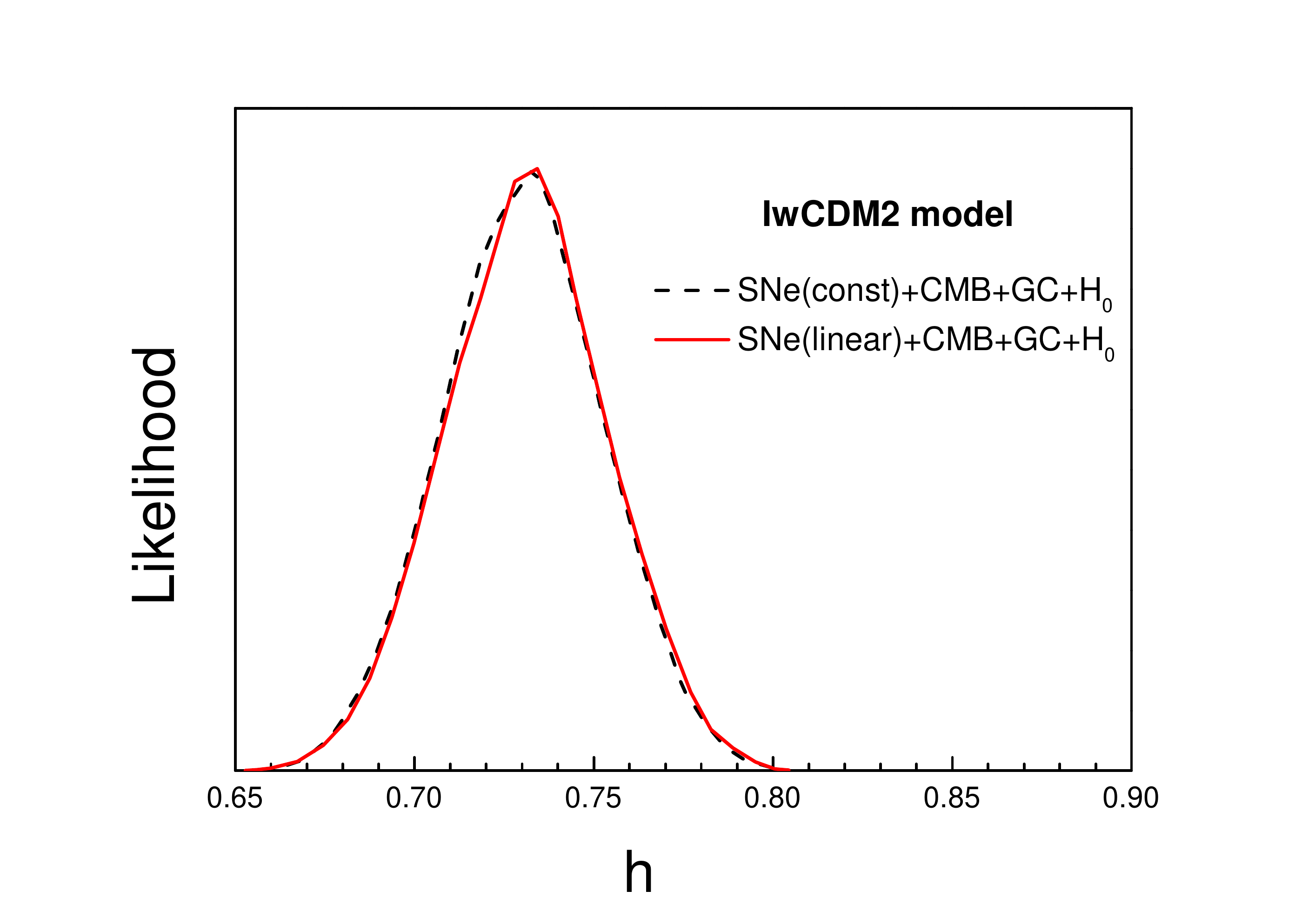}\hskip.4cm
\includegraphics[width=0.35\textwidth]{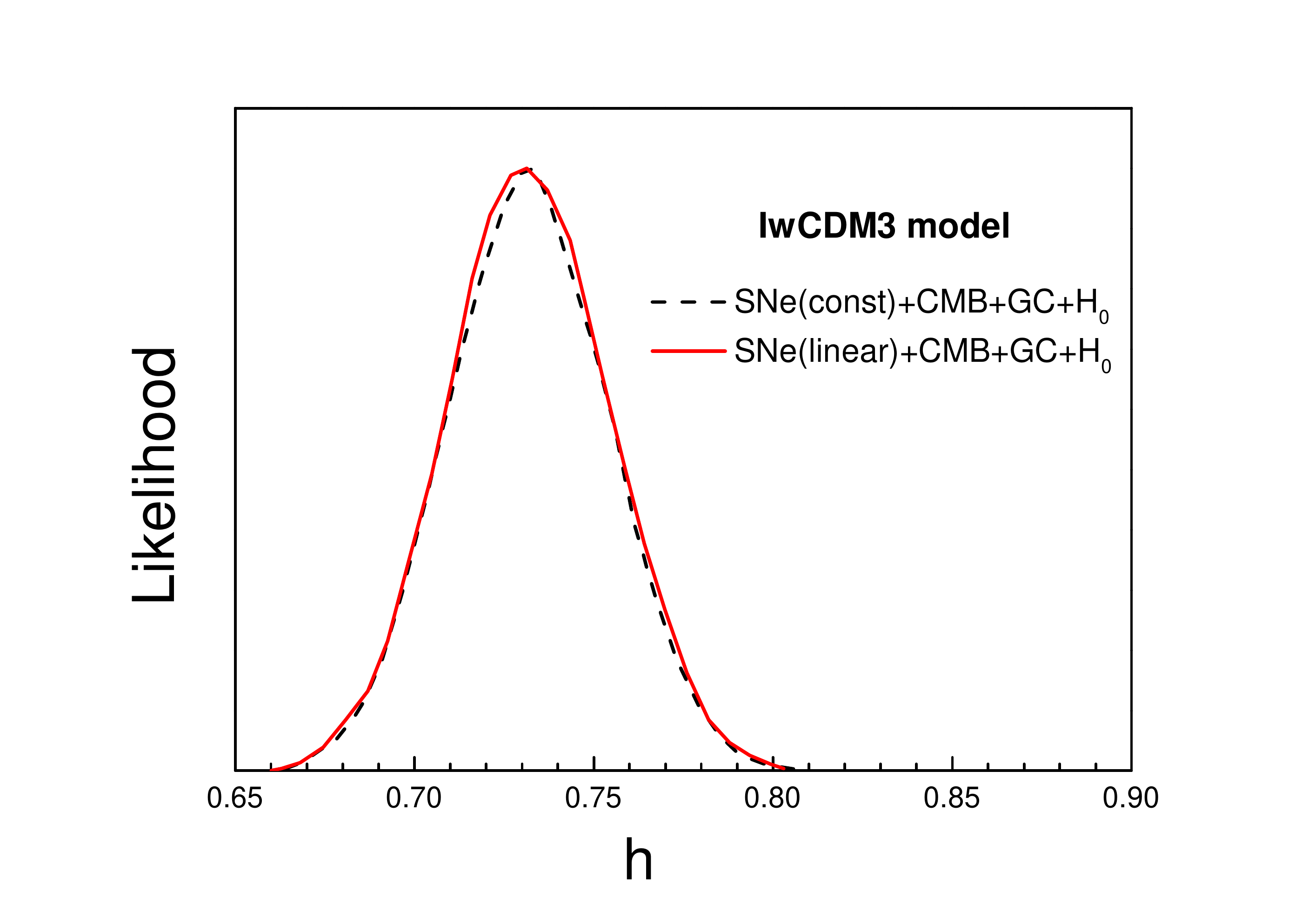}
\caption{\label{fig4}\footnotesize%
The 1D marginalized probability distributions of $h$, given by the SNe+CMB+GC+$H_0$ data,
for the $w$CDM model and the three IDE models.
Both the results of constant $\beta$ (dashed black lines) and linear $\beta(z)$ (solid red lines) cases are presented.}
\end{figure}

In Fig. \ref{fig4},
we plot the 1D marginalized probability distributions of $h$, for all the models considered in this paper.
It can be seen that, for the $w$CDM model, varying $\beta$ yields a smaller $h$;
this result is consistent with the noninteracting cases \cite{WangNew,WangNew2}.
However, for all the IDE models,
the 1D distribution results of $h$ of the linear $\beta$ case are almost same with those of the constant $\beta$ case.
In other words, once considering the interaction between dark sectors,
varying $\beta$ will not change the fitting results of $h$.
This result is quite different from the results of Fig. \ref{fig2} and Fig. \ref{fig3},
showing that there is a degeneracy between $h$ and $\gamma$.

\begin{figure}
\includegraphics[width=0.35\textwidth]{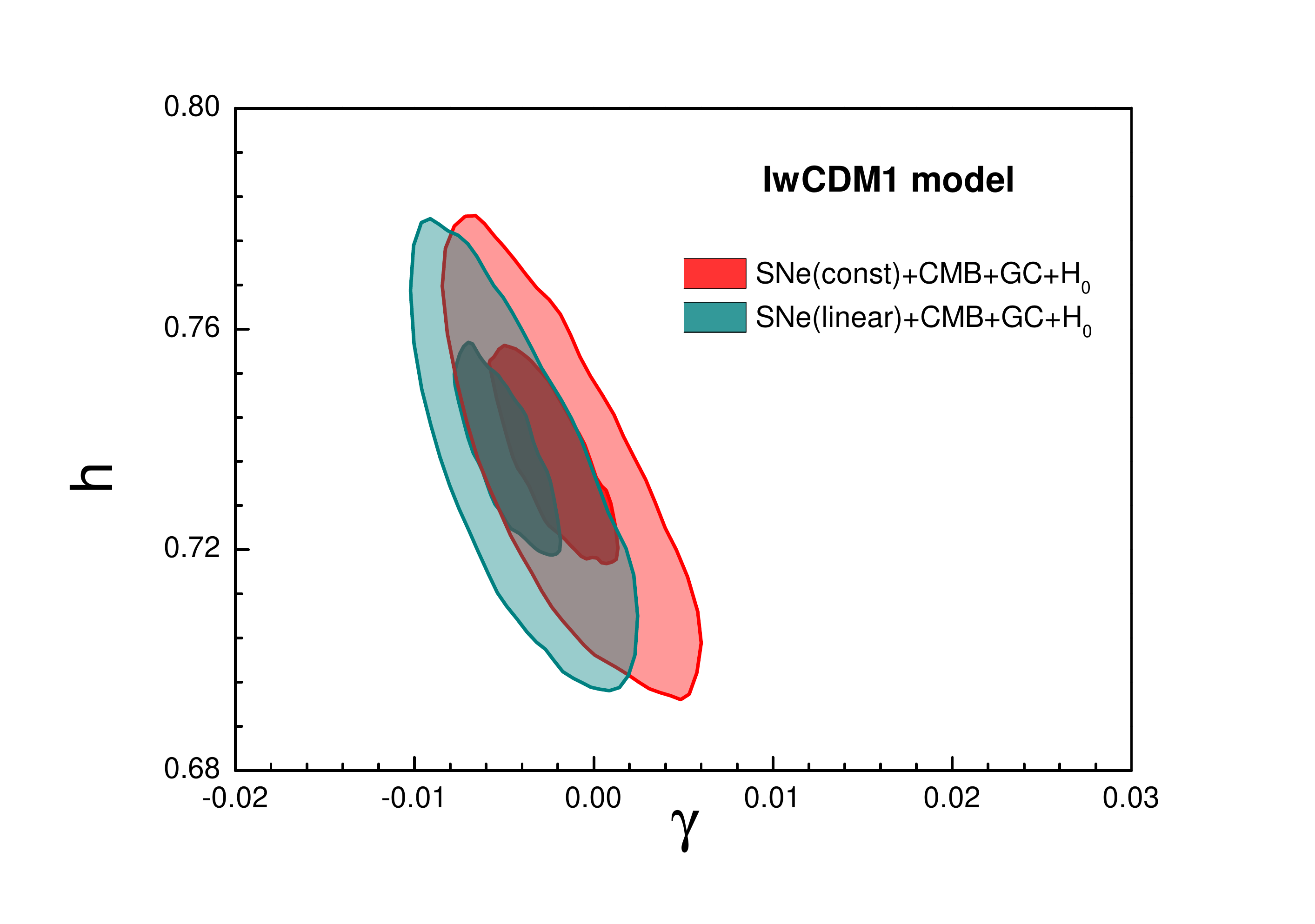}\hskip.4cm
\includegraphics[width=0.35\textwidth]{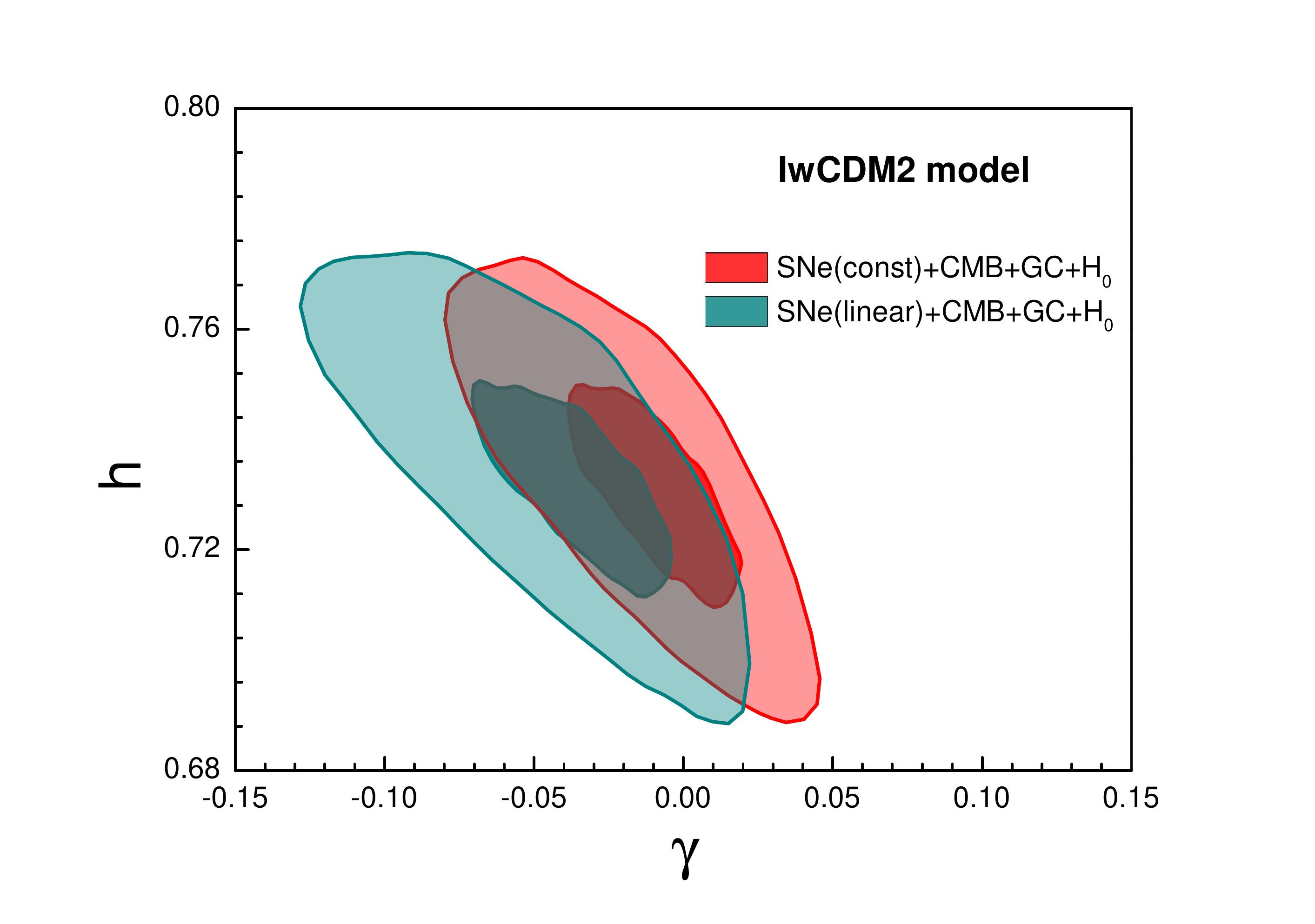}\vskip.4cm
\includegraphics[width=0.35\textwidth]{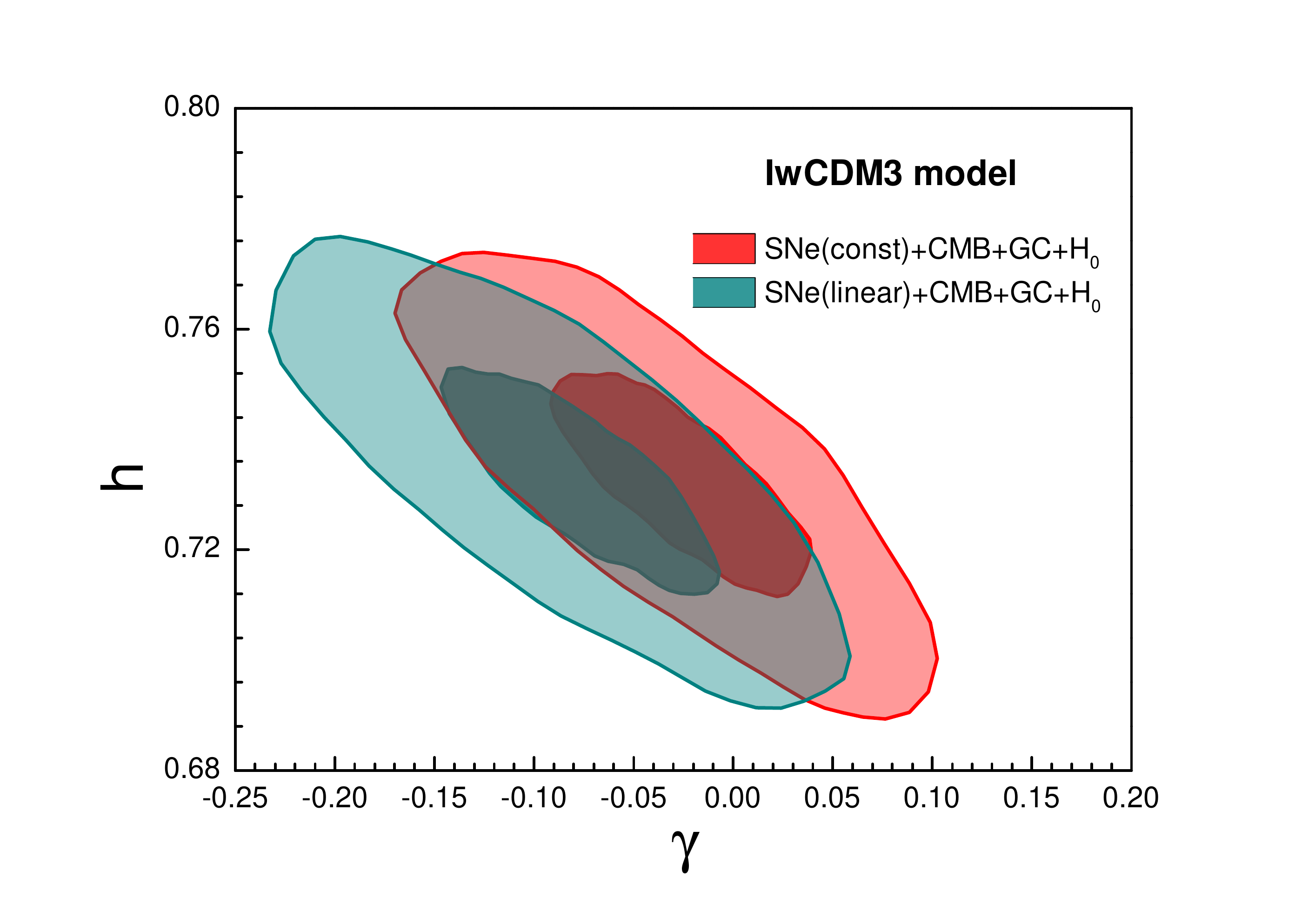}
\caption{\label{fig5}\footnotesize%
The 1$\sigma$ and 2$\sigma$ confidence contours for $\{\gamma,h\}$, for the three IDE models.
Both the results of constant $\beta$ (red regions) and linear $\beta(z)$ (dark cyan regions) cases are presented.}
\end{figure}

Next, we turn to the constraints on interaction parameter $\gamma$.
In Fig. \ref{fig5},
we plot 1$\sigma$ and 2$\sigma$ confidence contours for $\{\gamma,h\}$, for all the IDE models.
Again, one can see that varying $\beta$ has no impact on $h$.
In contrary, varying $\beta$ yields a smaller $\gamma$:
for the constant $\beta$ case, the best-fit results are $\gamma=-0.0028, -0.0105$, and $-0.0198$,
for the I$w$CDM1 model, the I$w$CDM2 model, and the I$w$CDM3 model, respectively;
while for the linear $\beta(z)$ case, the best-fit results are $\gamma=-0.0053, -0.0322$, and $-0.0732$,
for the I$w$CDM1 model, the I$w$CDM2 model, and the I$w$CDM3 model, respectively.
In other words, $\gamma<0$ is slightly more favored in the linear $\beta(z)$ case.
This means that energy will transfer from dark matter to dark energy.
In addition, we find that $\gamma$ and $h$ are anti-correlated,
showing that there is a degeneracy between $h$ and $\gamma$.

\begin{figure}
\psfig{file=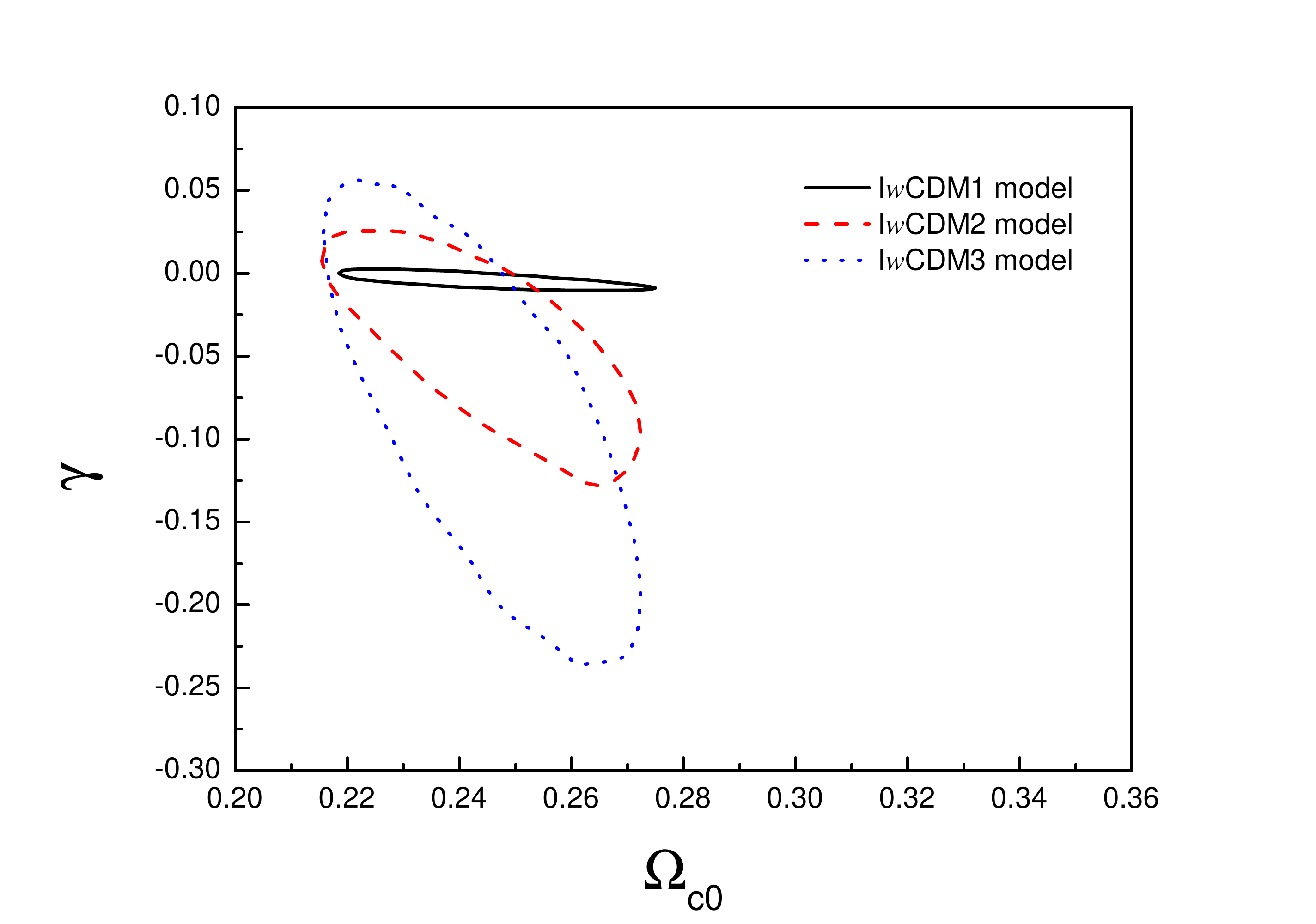,width=3.5in}\\
\caption{\label{fig6}\footnotesize%
The 2$\sigma$ confidence contours for $\{\Omega_{c0},\gamma\}$,
based on the linear $\beta(z)$ case,
for the I$w$CDM1 model (solid black line), the I$w$CDM2 model (dashed red line), and the I$w$CDM3 model (dotted blue line).
}
\end{figure}

In Fig. \ref{fig6}, to make a visual comparison among three interaction forms,
we plot the 2$\sigma$ confidence contours for $\{\Omega_{c0},\gamma\}$,
based on the linear $\beta(z)$ case, for all the IDE models.
From this figure one can see that, $\gamma$ is tightly constrained in the I$w$CDM1 model;
in contrary, $\gamma$ cannot be well constrained in the I$w$CDM2 and I$w$CDM3 models.
This result is consistent with the result of \cite{Geng}, in which only the constant $\beta$ case was considered.

\section{Discussion and summary}

In recent years, more and more SNe Ia have been discovered,
and the systematic errors of SNe Ia have drawn more and more attentions.
One of the most important systematic uncertainties for SNe Ia is the potential SN evolution.
The hints for the evolution of $\beta$ have been found \cite{Astier06,Kessler09,Marriner11,Scolnic1,Scolnic2}.
For examples, Mohlabeng and Ralston \cite{Mohlabeng} studied the case of Union2.1
and found that $\beta$ deviates from a constant at 7$\sigma$ CL.
In \cite{WangWang}, Wang \& Wang found that, for the SNLS3 data,
$\beta$ increases significantly with $z$ at the 6$\sigma$ CL;
moreover, they proved that this conclusion is insensitive to the lightcurve fitter models,
or the functional form of $\beta(z)$ assumed \cite{WangWang}.

It is clear that a time-varying $\beta$ will have significant impact on parameter estimation.
Adopting a constant $\alpha$ and a linear $\beta(z) = \beta_{0} + \beta_{1} z$,
Wang, Li \& Zhang \cite{WangNew} explored this issue by considering the $\Lambda$CDM model, the $w$CDM model, and the CPL model.
Soon after, Wang, Geng, Hu \& Zhang \cite{WangNew2} studied this issue in the frame of HDE model,
which is a physically plausible DE candidate based on the holographic principle.
It is found that, for all these DE models, $\beta$ deviates from a constant at 6$\sigma$ CL;
in addition, considering the evolution of $\beta$ is helpful in reducing the tension between SN and other cosmological observations.
It should be pointed out that,
in principle there is always an important possibility that DE directly interacts with CDM.
This factor was not considered in Refs. \cite{WangNew} and  \cite{WangNew2}.

In this paper, we extend the corresponding discussions to the case of IDE model.
To perform the cosmology-fits, the $w$CDM model is adopted.
Moreover, three kinds of interaction forms are considered:
$Q_{1}=3\gamma H\rho_{c}$, $Q_{2}=3\gamma H\rho_{de}$, and $Q_{3}=3\gamma H\frac{\rho_{c}\rho_{de}}{\rho_{c}+\rho_{de}}$.
In addition to the SNLS3 SN data,
we also use the Planck distance priors data,
the GC data extracted from SDSS DR7 and BOSS,
as well as the direct measurement of Hubble constant from the HST observation.

We further confirm the redshift-evolution of $\beta$ for the SNLS3 data:
for all the IDE models, adding a parameter of $\beta$ can reduce $\chi^2$ by $\sim$ 34,
indicating that $\beta_1 = 0$ is ruled out at 5.8$\sigma$ CL.
In addition, we find that the 1$\sigma$ regions of $\beta(z)$ of all these models are almost overlapping,
showing that the evolution of $\beta$ is insensitive to the interaction between dark sectors.
These results further verify the importance of considering the evolution of $\beta$ in the cosmology-fits.

Furthermore, we find that a time-varying $\beta$ has significant effects on the results of parameter estimation:
for all the models considered in this paper,
varying $\beta$ yields a larger $\Omega_{c0}$ and a larger $w$;
on the other side, varying $\beta$ yields a smaller $h$ for the $w$CDM model,
while varying $\beta$ has no influence on $h$ for the three IDE models.
Moreover, we find that $\gamma$ and $h$ are anti-correlated,
showing that there is a degeneracy between $h$ and $\gamma$.
In addition, we find that $\gamma$ is tightly constrained in the I$w$CDM1 model,
but cannot be well constrained in the I$w$CDM2 and I$w$CDM3 models.

In all, these results show that the evolution of $\beta$ is insensitive to the interaction between dark sectors,
and highlight the importance of considering $\beta$'s evolution in the cosmology fits.

So far, only the effects of varying $\beta$ on DE models are considered.
It is of great interest to study the effects of varying $\beta$ on parameter estimation in MG models.
In addition, some other factors, such as the evolution of $\sigma_{int}$ \cite{Kim2011},
may also cause the systematic uncertainties of SNe Ia.
These issues will be studied in future works.

\begin{acknowledgments}
We thank the referee for valuable suggestions, which help us to improve this work. 
We also thank Dr. Daniel Scolnic for helpful discussions.
We are grateful to Dr. Alex Conley for providing us with the SNLS3 covariance matrices that allow redshift-dependent $\beta$. 
We acknowledge the use of CosmoMC.
SW is supported by the Fundamental Research Funds for the Central Universities under Grant No. N130305007.
XZ is supported by the National Natural Science Foundation of China under Grant No. 11175042
and the Fundamental Research Funds for the Central Universities under Grant No. N120505003.
\end{acknowledgments}


\begin{thebibliography}{}


\bibitem{Riess98}
A. G. Riess {\it et al.}, AJ. {\bf 116}, 1009 (1998);
S. Perlmutter {\it et al.}, ApJ. {\bf 517}, 565 (1999).

\bibitem{spergel03}
D. N. Spergel {\it et al.}, ApJS {\bf 148}, 175 (2003);
C. L. Bennet {\it et al.}, ApJS. {\bf 148}, 1 (2003);
D. N. Spergel {\it et al.}, ApJS {\bf 170}, 377 (2007);
L. Page {\it et al.}, ApJS {\bf 170}, 335 (2007);
G. Hinshaw {\it et al.}, ApJS {\bf 170}, 263 (2007).

\bibitem{Tegmark04}
M. Tegmark {\it et al.}, Phys. Rev. D {\bf 69}, 103501 (2004);
ApJ {\bf 606}, 702 (2004); Phys. Rev. D {\bf 74}, 123507 (2006).


\bibitem{Komatsu09}
E. Komatsu {\it et al.}, ApJS. {\bf 180}, 330 (2009);
E. Komatsu {\it et al.}, ApJS. {\bf 192}, 18 (2011).


\bibitem{Percival10}
W. J. Percival {\it et al.}, MNRAS {\bf 401}, 2148 (2010);
A. G. Sanchez, {\it et al.}, arXiv:1203.6616, MNRAS accepted.

\bibitem{Drinkwater10}
M. Drinkwater {\it et al.}, MNRAS {\bf 401}, 1429 (2010);
C. Blake {\it et al.}, arXiv:1108.2635, MNRAS accepted.

\bibitem{Riess11}
A. G. Riess {\it et al.}, ApJ. {\bf 730}, 119 (2011).


\bibitem{quint}
P. J. E. Peebles and B. Ratra,  ApJ {\bf 325}, L17 (1988);
C. Wetterich, Nucl. Phys. B {\bf 302}, 668 (1988);
R. R. Caldwell, R. Dave and P. J. Steinhardt, Phys. Rev. Lett. {\bf 80}, 1582 (1998);
I. Zlatev, L. Wang and P. J. Steinhardt, Phys. Rev. Lett. {\bf 82}, 896 (1999).


\bibitem{phantom}
R. R. Caldwell, Phys. Lett. B {\bf 545}, 23 (2002);
S. M. Carroll, M. Hoffman and M. Trodden, Phys. Rev. D {\bf 68}, 023509 (2003);
R. R. Caldwell, M. Kamionkowski and N. N. Weinberg, Phys. Rev. Lett. {\bf 91}, 071301 (2003);
X. Zhang, Eur. Phys. J. C {\bf 59}, 755 (2009);
X. Zhang, Eur. Phys. J. C {\bf 60}, 661 (2009);
X. D. Li et al., Sci. China Phys. Mech. Astron. {\bf 55}, 1330 (2012).



\bibitem{k}
C. Armendariz-Picon, T. Damour and V. Mukhanov, Phys. Lett. B {\bf 458}, 209 (1999);
C. Armendariz-Picon, V. Mukhanov and P. J. Steinhardt, Phys. Rev. D {\bf 63}, 103510 (2001);
T. Chiba, T. Okabe and M. Yamaguchi, Phys. Rev. D {\bf 62}, 023511 (2000).

\bibitem{Chaplygin}
A. Y. Kamenshchik, U. Moschella and V. Pasquier, Phys. Lett. B {\bf 511}, 265 (2001);
M. C. Bento, O. Bertolami and A. A. Sen, Phys. Rev. D {\bf 66}, 043507 (2002).

\bibitem{ngcg}
X. Zhang, F. Q. Wu and J. Zhang, JCAP {\bf 01}, 003 (2006);
K. Liao, Y. Pan and Z. H. Zhu, Res. Astron. Astrophys. {\bf 13}, 159 (2013).

\bibitem{tachyonic}
T. Padmanabhan, Phys. Rev. D {\bf 66}, 021301 (2002);
J. S. Bagla, H. K. Jassal, and T. Padmanabhan, Phys. Rev. D {\bf 67}, 063504 (2003).



\bibitem{HDE}
M. Li, Phys. Lett. B {\bf 603}, 1 (2004);
Q. G. Huang and M. Li, JCAP {\bf 08}, 013 (2004).
X. Zhang and F. Q. Wu, Phys. Rev. D {\bf 72}, 043524 (2005);
Z. Chang, F. Q. Wu and X. Zhang, Phys. Lett. B {\bf 633}, 14 (2006);
X. Zhang and F. Q. Wu, Phys. Rev. D {\bf 76}, 023502 (2007);
J.~-F.~Zhang, X.~Zhang and H.~-Y.~Liu,
  %``Holographic dark energy in a cyclic universe,''
  Eur.\ Phys.\ J.\ C {\bf 52}, 693 (2007);
%  [arXiv:0708.3121 [hep-th]].
M. Li, C. S. Lin and Y. Wang, JCAP {\bf 05}, 023 (2008);
M. Li, X. D. Li, S. Wang and X. Zhang, JCAP {\bf 06}, 036 (2009);
%M. Li {\it et al.}, JCAP {\bf 12}, 014 (2009);
X. Zhang, Phys.\ Lett.\ B {\bf 683}, 81 (2010);
Y. H. Li, S. Wang, X. D. Li and X. Zhang, JCAP {\bf 02}, 033 (2013).
%M. Li, X. D. Li, Y. Z. Ma, X. Zhang and Z. H. Zhang, JCAP {\bf 09}, 021 (2013).


\bibitem{hessence}
H. Wei, R. G. Cai, and D. F. Zeng, Class. Quant. Grav. {\bf 22}, 3189 (2005);
H. Wei, and R. G. Cai, Phys. Rev. D {\bf 72}, 123507 (2005);
H. Wei, N. Tang, and S. N. Zhang, Phys. Rev. D{\bf 75}, 043009 (2007).


\bibitem{YMC}
W. Zhao and Y. Zhang, Class. Quant. Grav. {\bf 23}, 3405 (2006);
T. Y. Xia and Y. Zhang, Phys. Lett. B {\bf 656}, 19 (2007);
S. Wang, Y. Zhang and T. Y. Xia, JCAP {\bf 10}, 037 (2008);
S. Wang and Y. Zhang, Phys. Lett. B {\bf 669}, 201 (2008).

\bibitem{hscalar}
X. Zhang, Phys. Lett. B {\bf 648}, 1 (2007);
X. Zhang, Phys. Rev. D {\bf 74}, 103505 (2006);
J. Zhang, X. Zhang and H. Liu, Phys. Lett. B {\bf 651}, 84 (2007);
J. Zhang, X. Zhang and H. Liu, Eur. Phys. J. C {\bf 54}, 303 (2008);
X. Zhang, Phys. Rev. D {\bf 79}, 103509 (2009).



\bibitem{cq}
D. Comelli, M. Pietroni and A. Riotto, Phys. Lett. B {\bf 571}, 115 (2003);
X. Zhang, Mod. Phys. Lett. A {\bf 20}, 2575 (2005);
X. Zhang, Phys. Lett. B {\bf 611}, 1 (2005).


\bibitem{others1}
J. A. Frieman, C. T. Hill, A. Stebbins, and I. Waga, Phys. Rev. Lett. {\bf 75}, 2077 (1995);
M. Chevallier and D. Polarski, Int. J. Mod. Phys. D {\bf 10}, 213 (2001);
E. V. Linder, Phys. Rev. Lett. {\bf 90} 091301 (2003);
D. Huterer and G. Starkman, Phys. Rev. Lett. {\bf 90}, 031301 (2003);
D. Huterer and A. Cooray, Phys. Rev. D {\bf 71}, 023506 (2005).


\bibitem{others2}
Y. Wang and M. Tegmark, Phys. Rev. Lett. {\bf 92}, 241302 (2004);
Y. Wang and M. Tegmark, Phys. Rev. D {\bf 71}, 103513 (2005);
Y. Wang, and K. Freese, Phys. Lett. B {\bf 632}, 449 (2006);
Y. Wang and P. Mukherjee, ApJ. {\bf 650}, 1 (2006);
Y. Wang and P. Mukherjee, Phys. Rev. D {\bf 76}, 103533 (2007);
Y. Wang,Phys. Rev. D {\bf 78}, 123532 (2008).


\bibitem{others3}
U. Alam, V. Sahni, T. D. Saini and A. A. Starobinsky, Mon. Not. Roy. Astron. Soc. {\bf 344}, 1057 (2003);
U. Alam, V. Sahni, T. D. Saini and A. A. Starobinsky, Mon. Not. Roy. Astron. Soc. {\bf 354}, 275 (2004);
A. Shafieloo, U. Alam, V. Sahni and A. A. Starobinsky, Mon. Not. Roy. Astron. Soc. {\bf 366}, 1081 (2006);
U. Alam, V. Sahni and A. A. Starobinsky, JCAP {\bf 02}, 011 (2007);
V. Sahni, A. Shafieloo and A. A. Starobinsky, Phys. Rev. D {\bf 78}, 103502, (2008);
A. Shafieloo, V. Sahni and A. A. Starobinsky, Phys. Rev. D {\bf 80}, 101301(R) (2009);
A. Shafieloo, V. Sahni and A. A. Starobinsky, Phys. Rev. D {\bf 86}, 103527 (2012).


\bibitem{others4}
J. F. Zhang, X. Zhang and H. Y. Liu, Mod. Phys. Lett. A {\bf 23}, 139 (2008);
Q. G. Huang, M. Li, X. D. Li and S. Wang, Phys. Rev. D {\bf 80}, 083515 (2009);
S. Wang, X. D. Li and M. Li, Phys. Rev. D {\bf 82}, 103006 (2010);
M. Li, X. D. Li and X. Zhang, Sci. China Phys. Mech. Astron. {\bf 53}, 1631 (2010);
S. Wang, X. D. Li and M. Li, Phys. Rev. D {\bf 83}, 023010 (2011);
Y. H. Li and X. Zhang, Eur. Phys. J. C {\bf 71}, 1700 (2011);
X. D. Li {\it et al.}, JCAP {\bf 07}, 011 (2011);
J. Z. Ma and X. Zhang, Phys. Lett. B {\bf 699}, 233 (2011);
H. Li and X. Zhang, Phys. Lett. B {\bf 713}, 160 (2012).



\bibitem{SH}
V. Sahni and S. Habib, Phys. Rev. Lett. {\bf 81}, 1766 (1998).

\bibitem{PR}
L. Parker and A. Raval, Phys. Rev. D {\bf 60}, 063512 (1999).

\bibitem{DGP}
G. Dvali, G. Gabadadze and M. Porrati, Phys. Lett. B {\bf 485}, 208 (2000).

\bibitem{GB}
S. Nojiri, S. D. Odintsov, and M. Sasaki, Phys. Rev. D {\bf 71}, 123509 (2005).

\bibitem{Galileon}
A. Nicolis, R. Rattazzi, and E. Trincherini, Phys. Rev. D {\bf 79}, 064036 (2009).

\bibitem{FR}
W. Hu and I. Sawicki, Phys. Rev. D {\bf 76}, 064004 (2007);
A. A. Starobinsky, J. Exp. Theor. Phys. Lett. {\bf 86}, (2007) 157.

\bibitem{FT}
G. R. Bengochea and R. Ferraro, Phys. Rev. D {\bf 79}, 124019 (2009);
E. V. Linder, Phys. Rev. D {\bf 81}, (2010) 127301.

\bibitem{FRT}
T. Harko, F. S. N. Lobo, S. Nojiri and S. D. Odintsov, Phys. Rev. D {\bf 84}, 024020 (2011).


\bibitem{CST}
E. J. Copeland, M. Sami and S. Tsujikawa, Int. J. Mod. Phys. D {\bf 15}, 1753 (2006).

\bibitem{FTH}
J. Frieman, M. Turner and D. Huterer, Ann. Rev. Astron. Astrophys {\bf 46}, 385 (2008).


\bibitem{Linder}
E. V. Linder, Rept. Prog. Phys. {\bf 71}, 056901 (2008).

\bibitem{CK}
R. R. Caldwell and M. Kamionkowski, Ann. Rev. Nucl. Part. Sci. {\bf 59}, 397 (2009).

\bibitem{Uzan}
J.-P. Uzan, arxiv:0908.2243.

\bibitem{Tsujikawa}
S. Tsujikawa, arXiv:1004.1493.

\bibitem{NO}
S. Nojiri and S. D. Odintsov, Phys. Rept. {\bf 505}, 59 (2011).

\bibitem{LLWW}
M. Li, X. D. Li, S. Wang and Y. Wang, Commun. Theor. Phys. {\bf 56}, 525 (2011).

\bibitem{CFPS}
T. Clifton, P. G. Ferreira, A. Padilla and C. Skordis, Phys. Rept. {\bf 513}, 1 (2012).

\bibitem{YWBook}
Y. Wang, {\it Dark Energy}, Wiley-VCH (2010).


\bibitem{Union}
M. Kowalski, {\it et al.}, ApJ. {\bf 686}, 749 (2008).

\bibitem{Constitution}
M. Hicken, {\it et al.}, ApJ. {\bf 700}, 1097 (2009);
M. Hicken, {\it et al.}, ApJ. {\bf 700}, 331 (2009).

\bibitem{Union2}
R. Amanullah, {\it et al.}, ApJ. {\bf 716}, 712 (2010).

\bibitem{Union2.1}
N. Suzuki, {\it et al.}, ApJ {\bf 746}, 85 (2012).


\bibitem{Guy10}
J. Guy, {\it et al.}, A\&A, {\bf 523}, 7 (2010).

\bibitem{Conley11}
A. Conley, {\it et al.}, ApJS. 192 1 (2011). % -- {\bf C11}

\bibitem{Sullivan11}
M. Sullivan, {\it et al.}, arXiv:1104.1444.


\bibitem{Astier06}
Astier, {\it et al.}, Astron. Astrophys. {\bf 447}, 31 (2006).

\bibitem{Kessler09}
R. Kessler, {\it et al.}, ApJS. {\bf 185}, 32 (2009).

\bibitem{Marriner11}
Marriner, {\it et al.}, arXiv:1107.4631.


\bibitem{Scolnic1}
D. Scolnic, {\it et al.},  arXiv:1306.4050, ApJ in press.

\bibitem{Scolnic2}
D. Scolnic, {\it et al.},  arXiv:1310.3824.


\bibitem{Mohlabeng}
G.Mohlabeng and J. Ralston, arXiv:1303.0580.

\bibitem{WangWang}
S. Wang and Y. Wang, Phys. Rev. D {\bf 88}, 043511 (2013).


\bibitem{WangNew}
S. Wang, Y. H. Li and X. Zhang, Phys. Rev. D {\bf 89}, 063524 (2014).


\bibitem{WangNew2}
S. Wang, J. J. Geng, Y. L. Hu and X. Zhang, arXiv:1312.0184.

\bibitem{WangWangCMB}
Y. Wang and S. Wang, Phys. Rev. D {\bf 88}, 043522 (2013).

\bibitem{ChuangWang12}
C. H. Chuang and Y. Wang, MNRAS, {\bf 426}, 226 (2012).


\bibitem{Chuang13}
C. H. Chuang, {\it et al.}, arXiv:1303.4486.


\bibitem{IDEPaper1}
G. R. Farrar and P. J. E. Peebles, Astrophys. J. {\bf 604}, 1 (2004);
G. Olivares, F. Atrio-Barandela, and D. Pavon, Phys. Rev. D {\bf 71}, 063523 (2005);
T. Koivisto, Phys. Rev. D {\bf 72}, 043516 (2005);
H. M. Sadjadi and M. Alimohammadi, Phys. Rev. D {\bf 74}, 103007 (2006);
Z. K. Guo, N. Ohta, and S. Tsujikawa, Phys. Rev. D 76, 023508 (2007).


\bibitem{IDEPaper2}
C. G. Boehmer, G. Caldera-Cabral, R. Lazkoz, and R. Maartens, Phys. Rev. D {\bf 78}, 023505 (2008);
M. Quartin, M. O. Calvao, S. E. Joras, R. R. R. Reis, and I. Waga, JCAP {\bf 05}, 007 (2008);
J. Valiviita, E. Majerotto, and R. Maartens, J. Cosmol. Astropart. Phys. {\bf 07} 020 (2008);
R. Bean, E. E. Flanagan, I. Laszlo, and M. Trodden, Phys. Rev. D {\bf 78}, 123514 (2008);
S. Chongchitnan, Phys. Rev. D {\bf 79}, 043522 (2009);
B. M. Jackson, A. Taylor, and A. Berera, Phys. Rev. D {\bf 79}, 043526 (2009).


\bibitem{IDEPaper3}
J. Zhang, H. Liu, and X. Zhang, Phys. Lett. B {\bf 659}, 26 (2008);
M.~Li, X.~-D.~Li, S.~Wang, Y.~Wang and X.~Zhang,
  %``Probing interaction and spatial curvature in the holographic dark energy model,''
  JCAP {\bf 0912}, 014 (2009);
 % [arXiv:0910.3855 [astro-ph.CO]]. 
L.~Zhang, J.~Cui, J.~Zhang and X.~Zhang,
  %``Interacting model of new agegraphic dark energy: Cosmological evolution and statefinder diagnostic,''
  Int.\ J.\ Mod.\ Phys.\ D {\bf 19}, 21 (2010);
%  [arXiv:0911.2838 [astro-ph.CO]].
J.~Cui and X.~Zhang,
  %``Cosmic age problem revisited in the holographic dark energy model,''
  Phys.\ Lett.\ B {\bf 690}, 233 (2010);
%  [arXiv:1005.3587 [astro-ph.CO]].
%R. G. Cai and Q. Su, Phys. Rev. D {\bf 81}, 103514 (2010);
%Y. H. Li and X. Zhang, Eur. Phys. J. C {\bf 71}, 1700 (2011);
Y.~Li, J.~Ma, J.~Cui, Z.~Wang and X.~Zhang,
  %``Interacting model of new agegraphic dark energy: observational constraints and age problem,''
  Sci.\ China Phys.\ Mech.\ Astron. {\bf 54}, 1367 (2011);
%  [arXiv:1011.6122 [astro-ph.CO]].
T.~-F.~Fu, J.~-F.~Zhang, J.~-Q.~Chen and X.~Zhang,
  %``Holographic Ricci dark energy: Interacting model and cosmological constraints,''
  Eur.\ Phys.\ J.\ C {\bf 72}, 1932 (2012);
%  [arXiv:1112.2350 [astro-ph.CO]].
Z.~Zhang, S.~Li, X.~-D.~Li, X.~Zhang and M.~Li,
  %``Revisit of the Interaction between Holographic Dark Energy and Dark Matter,''
  JCAP {\bf 1206}, 009 (2012);
%  [arXiv:1204.6135 [astro-ph.CO]].
T. Clemson, K. Koyama, G. B. Zhao, R. Maartens, and J. Valiviita, Phys. Rev. D {\bf 85}, 043007 (2012); 
J.~Zhang, L.~Zhao and X.~Zhang,
  %``Revisiting the interacting model of new agegraphic dark energy,''
  Sci.\ China Phys.\ Mech.\ Astron.\  {\bf 57}, 387 (2014).
%  [arXiv:1306.1289 [astro-ph.CO]].



\bibitem{IDEPaper4}
J. H. He and B. Wang, JCAP. {\bf 06}, 010 (2008);
J. H. He, B. Wang, and P. Zhang, Phys. Rev. D {\bf 80}, 063530 (2009);
J. H. He, B. Wang, E. Abdalla, and D. Pavon, JCAP. {\bf 12}, 022 (2010);
X. D. Xu, B. Wang, and E. Abdalla, Phys. Rev. D {\bf 85}, 083513 (2012);
X. D. Xu, B. Wang, P. Zhang, and F. Atrio-Barandela, JCAP. {\bf 12}, 001 (2013).



\bibitem{IDEP2}
Y. H. Li and X. Zhang, Phys. Rev. D {\bf 89}, 083009 (2014).


\bibitem{Kim2011}
A. Kim, arXiv:1101.3513;
J. Marriner, {\it et al.}, arXiv:1107.4631.


\bibitem{IDEP1}
Y. H. Li, J. F. Zhang and X. Zhang, arXiv:1404.5220.



\bibitem{COSMOMC}
A. Lewis and S. Bridle, Phys. Rev. D {\bf 66}, 103511 (2002).


\bibitem{Geng}
J. J. Geng, J. F. Zhang and X. Zhang, 
JCAP {\bf 07}, 006 (2014).
%arXiv:1404.5407.


\end{thebibliography}
\end{document}